\documentclass[11pt]{article}

\usepackage{color}

\newcommand{\eat}[1]{} 

\eat{}

\usepackage{acl}

\usepackage{times}
\usepackage{multirow}
\usepackage[table]{xcolor}
\usepackage{pifont}
\usepackage{latexsym}
\usepackage{todonotes}
\newcommand{\sys}{EAGR}

\usepackage[T1]{fontenc}

\usepackage[utf8]{inputenc}

\usepackage{microtype}

\usepackage{inconsolata}

\usepackage{graphicx}
\usepackage{tikz}
\usetikzlibrary{arrows.meta,positioning,calc}
\usepackage{algorithm}
\usepackage{algpseudocode}
\usepackage{tcolorbox}
\usepackage{amsmath}
\usepackage{array}
\usepackage{tabularx}
\usepackage{tcolorbox}
\usepackage{listings}
\usepackage{textcomp}
\tcbuselibrary{skins,listings}

\usepackage{tikz}
\usepackage{amsmath,amssymb}
\usepackage{booktabs}
\usetikzlibrary{arrows.meta,positioning,calc,fit,backgrounds}
 
\definecolor{eaInk}{HTML}{16202B}
\definecolor{eaSeed}{HTML}{2B6CB0}
\definecolor{eaRel}{HTML}{2F855A}
\definecolor{eaPrune}{HTML}{97A0AA}
\definecolor{eaOp}{HTML}{B45309}
\definecolor{eaPanel}{HTML}{F6F7F9}
\definecolor{eaLine}{HTML}{5A6472}
 
\tikzset{
  panel/.style    ={rounded corners=3pt, draw=eaInk!15, fill=eaPanel, line width=0.6pt},
  ptitle/.style   ={font=\sffamily\bfseries\scriptsize, text=eaInk, anchor=west},
  chead/.style    ={font=\sffamily\tiny, text=eaInk!75},
  lbl/.style      ={font=\sffamily\tiny, text=eaInk!75, inner sep=1pt},
  oplbl/.style    ={font=\sffamily\tiny, text=eaOp, inner sep=1pt},
  nbox/.style     ={rounded corners=2pt, draw=eaLine!55, fill=white, line width=0.6pt,
                    align=center, font=\sffamily\tiny, inner sep=2.5pt},
  psg/.style      ={nbox, text width=1.67cm, minimum width=1.85cm,
                    minimum height=1.05cm, align=left},
  relbox/.style   ={psg, draw=eaRel, fill=eaRel!7},
  prunebox/.style ={psg, draw=eaPrune, fill=eaPrune!12, text=eaInk!55},
  ghostbox/.style ={psg, draw=eaPrune, fill=white, text=eaInk!45,
                    dash pattern=on 1.6pt off 1.4pt},
  card/.style     ={rounded corners=2pt, draw=eaOp!75, fill=eaOp!6, line width=0.6pt,
                    align=center, font=\sffamily\tiny, inner sep=2.5pt,
                    text width=1.50cm, minimum width=1.68cm, minimum height=1.86cm},
  memo/.style     ={rounded corners=3pt, draw=eaOp!35, fill=eaOp!5, line width=0.6pt},
  chip/.style     ={rounded corners=2pt, draw=eaOp!45, fill=white, line width=0.5pt,
                    font=\sffamily\tiny, inner sep=2.5pt, align=left, text width=1.67cm, minimum width=1.85cm},
  flow/.style     ={-{Stealth[length=4pt,width=3pt]}, line width=0.75pt, draw=eaLine},
  thinflow/.style ={-{Stealth[length=3.4pt,width=2.6pt]}, line width=0.6pt, draw=eaLine!65},
  evflow/.style   ={-{Stealth[length=3.4pt,width=2.6pt]}, line width=0.65pt, draw=eaOp!85,
                    dash pattern=on 1.7pt off 1.3pt},
  gedge/.style    ={draw=eaInk!35, line width=0.55pt, dash pattern=on 1.4pt off 1.2pt},
}
\title{EAGR: Evidence-Guided Adaptive Graph Retrieval}

\author{
\textbf{Sonam Gupta}\textsuperscript{1}\thanks{Corresponding author.},
\textbf{Vignesh P}\textsuperscript{1},
\textbf{Vishwajeet Kumar}\textsuperscript{1},
\textbf{Adnan Qidwai}\textsuperscript{1},\\
\textbf{Anand Eswaran}\textsuperscript{1},
\textbf{Jaydeep Sen}\textsuperscript{1},
\textbf{Sachindra Joshi}\textsuperscript{1} \\[0.5em]
\textsuperscript{1}IBM \\
\small sonam.gupta7@ibm.com, vignesh.p@ibm.com, vishk024@in.ibm.com, adnan.qidwai1@ibm.com, \\ \small anand.eswaran@ibm.com, jaydesen@in.ibm.com, jsachind@in.ibm.com
}

\begin{document}
\maketitle
\begin{abstract}

\eat{
Retrieval-Augmented Generation (RAG) increasingly relies on graph-based retrieval to support multi-hop reasoning, but existing GraphRAG methods require explicit knowledge graph construction or expensive iterative query reformulation. We present Evidence-Guided Adaptive Graph Retrieval \sys{}, a lightweight retrieval framework that formulates retrieval as evidence-guided navigation over an implicit semantic graph induced by metadata pre-ingested into a standard VectorDB. Retrieval begins from dense-retrieval seed passages, after which an LLM iteratively conditions retrieval decisions on the original query and accumulated evidence to identify promising semantic directions, expand retrieval, and terminate once sufficient evidence has been collected. By shifting semantic reasoning to offline metadata enrichment, \sys{} eliminates the need for dedicated GraphDBs and iterative query rewriting while retaining adaptive graph exploration. Experiments against representative GraphRAG and adaptive retrieval baselines, including HippoRAG and MemGraphRAG, on popular multi-hop reasoning benchmarks demonstrate consistent average improvements of ~$5.5\%$ in retrieval quality and ~$11\%$ in end-to-end QA performance, compared to the latest graph rag baseline MemGraphRAG.
}

Retrieval-Augmented Generation (RAG) for multi-hop reasoning requires retrieval to adapt as intermediate evidence reveals new information needs. Existing approaches support such retrieval through explicit graph structures or iterative query reformulation, introducing graph-construction or query-time overhead. We present Evidence-Guided Adaptive Graph Retrieval (EAGR), a framework in which accumulated evidence dynamically determines which semantic directions to explore next, and which requires no dedicated graph database and no explicit graph materialization. EAGR operates over an implicit semantic graph induced by compact passage metadata pre-ingested into a standard VectorDB. Starting from dense-retrieval seed passages, an LLM iteratively assesses accumulated evidence, identifies promising semantic directions, expands the retrieval frontier through metadata-based neighborhood lookup, and terminates once sufficient evidence has been collected. By precomputing semantic metadata, EAGR enables evidence-conditioned graph exploration without materializing an explicit graph, maintaining a dedicated GraphDB, or iteratively rewriting the query. Across three widely used multi-hop QA benchmarks, EAGR achieves the highest LLM-based answer accuracy among evaluated methods, outperforming the vanilla RAG by 17\% and strongest graph-aware baseline by 4\% percentage points averaged across three popular benchmarks HotPotQA, 2WikiMultiHopQA, and MuSiQue.

\end{abstract}

\section{Introduction}

Retrieval-Augmented Generation (RAG) grounds Large Language Models (LLMs) in external knowledge~\cite{lewis2020rag}. Conventional RAG retrieves a fixed set of passages using dense similarity and delegates reasoning to the LLM~\cite{karpukhin2020dpr}. This works for localized evidence, but often struggles when an answer requires combining information across documents through dependent retrieval steps~\cite{xiong2021mdr,trivedi2023ircot}. Such questions require retrieval to evolve as evidence accumulates.

The full retrieval path is rarely apparent from the original query. For example, a question about a person's \emph{mother-in-law} may require first finding a passage about their \emph{spouse}, then using that evidence to retrieve information about the \emph{mother}. The next useful retrieval direction thus emerges from intermediate evidence, rather than from the query alone.

Recent work tackles multi-hop retrieval using graph-based structures, including hierarchical summary graphs, knowledge graphs, associative memory graphs, and relation-free passage graphs~\cite{graphrag,gutierrez2024hipporag,gutierrez2025hipporag2,memgraphrag,wang2024kgp,li2025gnnret}. These methods demonstrate the value of semantic connectivity, but commonly add a graph representation or indexing stage to the passage index. Another line of work treats retrieval over explicit knowledge graphs as a planning problem. Systems such as StructGPT, Think-on-Graph, and Plan-on-Graph use LLMs to explore graph neighborhoods and refine their decisions~\cite{structgpt,tog,pog}. They assume a symbolic graph whose entities and relations are known \emph{a priori}.

\begin{figure}
  \centering
  \includegraphics[width=\linewidth]{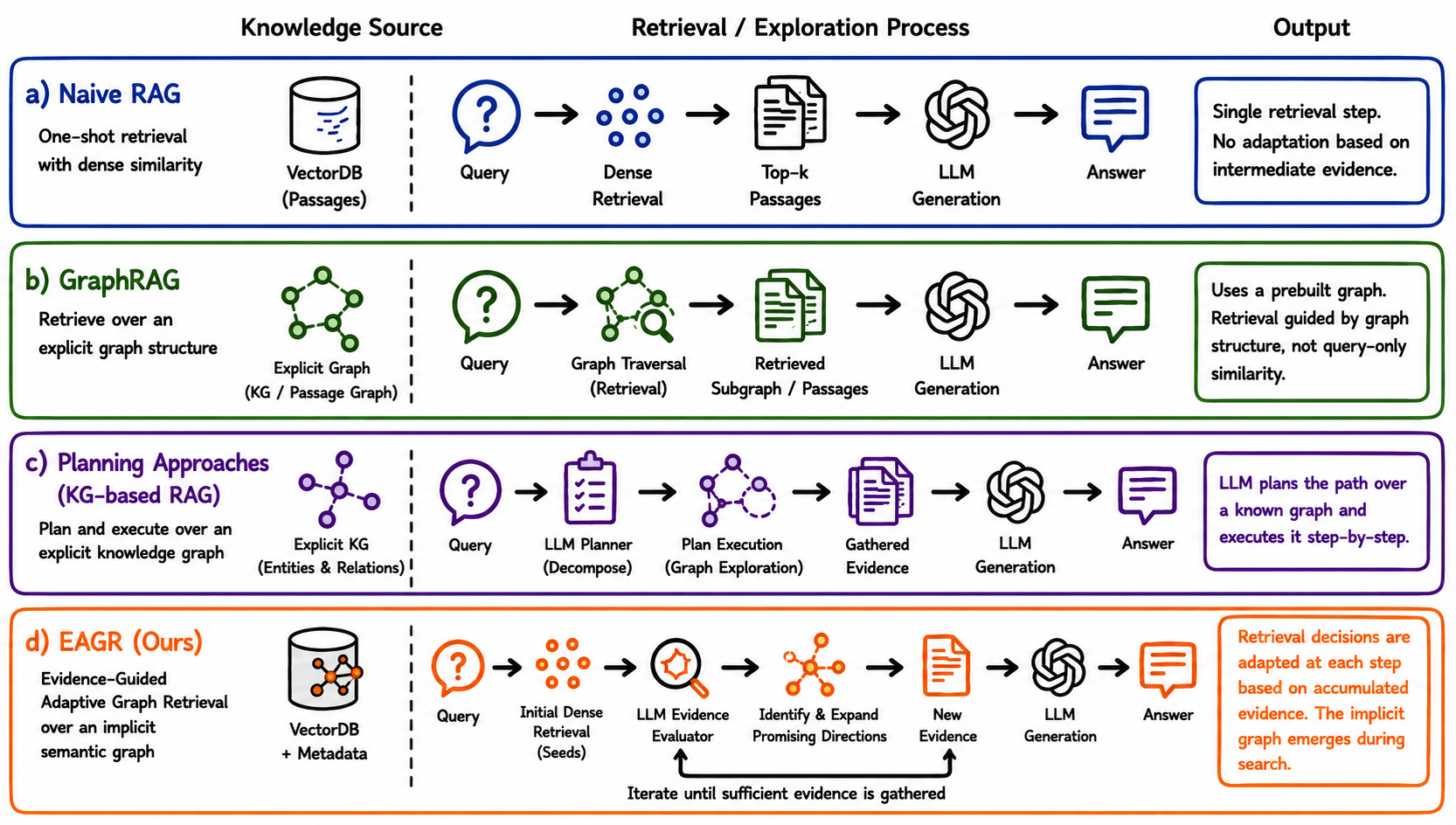}
  \caption{Comparison of various Retrieval Approaches}
  \label{fig:overview}
\end{figure}

We study a different setting: evidence-conditioned traversal over unstructured passages~\cite{li2026s2g}. Enterprise VectorDBs typically store passages and metadata, not explicit symbolic graphs. We derive passage connections from shared keyword and semantic descriptors in the existing index, then use metadata lookups to retrieve connected passages. This enables GraphRAG-style traversal without a separate graph store or a separately materialized graph. The central challenge is to discover which semantic directions to explore as evidence accumulates, rather than to execute a plan over a known graph.

We propose \textbf{Evidence-Guided Adaptive Graph Retrieval (\sys{})}, which views retrieval as an adaptive search process. Each passage is associated with LLM-generated semantic descriptors stored as metadata; shared descriptors induce implicit connections without explicit graph edges. Retrieval starts with dense-retrieved seed passages. At each iteration, an LLM assesses the accumulated evidence, selects promising directions, and expands the retrieval frontier through metadata-based neighborhood lookup. Unlike planning-based approaches, EAGR neither executes an explicit reasoning plan~\cite{pog,logicrag} nor repeatedly rewrites the query: its traversal adapts to the evidence gathered so far.

Our contributions are:

\begin{enumerate}
\item \textbf{Evidence-guided adaptive retrieval.} We formulate retrieval decisions around both the original query and accumulated evidence, enabling iterative LLM-guided exploration without repeated query rewriting.
\item \textbf{GraphRAG over a standard VectorDB.} We derive implicit passage connections from LLM-generated semantic descriptors stored as metadata, enabling neighborhood lookup without materialized graph edges or a separate graph store.
\item \textbf{EAGR traversal strategy.} We introduce an iterative procedure that selects semantic directions, expands the retrieval frontier, and stops when sufficient supporting context has been collected.
\end{enumerate}

Figure~\ref{fig:overview} contrasts EAGR with related retrieval approaches. The remainder of the paper is organized as follows. Section~\ref{sec:rel-work} reviews related work; Section~\ref{sec:method} presents EAGR; Sections~\ref{sec:experimentals} and~\ref{sec:results} describe the experimental setup and results; and Section~\ref{sec:concl} concludes with limitations.

\section{Related Work}
\label{sec:rel-work}

\paragraph{Adaptive retrieval.}
Dense retrievers rank passages once against the query~\cite{karpukhin2020dpr}.
Multi-hop methods instead condition later retrieval on earlier passages or the
model's reasoning state~\cite{xiong2021mdr,trivedi2023ircot,shao2023iterretgen,jiang2023flare},
and recent work uses intermediate knowledge, sufficiency judgements or
identified information gaps to steer subsequent
steps~\cite{zhuang2024efficientrag,jeong-etal-2024-adaptive,asai2024selfrag,su-etal-2024-dragin,fang2025kirag,li2026s2g,li2026grip}.
Most of these methods adapt by issuing new queries against the entire corpus
index. EAGR keeps the original query fixed and adapts through accumulated
evidence instead, restricting each step to the neighbourhood of the passages
already explored.

\paragraph{Graph-structured retrieval.}
Graph-based RAG organizes corpora into summary hierarchies, entity--relation
graphs or associative
networks~\cite{sarthi2024raptor,edge2024graphrag,gutierrez2024hipporag,gutierrez2025hipporag2,zhu-etal-2025-knowledge,guo2025lightrag,wu2026memgraphrag},
and lighter designs reduce construction cost with incidence links instead of
triples~\cite{huang2025ketrag,zhuang2026linearrag}. These structures expose
cross-passage connections, but they must be built and stored, and query-time
retrieval typically follows community hierarchies, graph scores such as
personalized PageRank, or fixed expansion rules. EAGR builds no graph and
replaces fixed rules with evidence-guided search.

\paragraph{LLM-guided graph traversal.}
Think-on-Graph~\cite{sun2024thinkongraph}, Plan-on-Graph~\cite{chen2024planongraph}
and ToG~2.0~\cite{ma2025thinkongraph2} let an LLM navigate symbolic knowledge
graphs, and GraphReader~\cite{li2024graphreader}, GeAR~\cite{shen2025gear} and
ReMindRAG~\cite{hu2025remindrag} navigate graphs extracted from text. All
assume a graph built in advance. Closest to EAGR are passage-level methods:
KGP links passages via shared keywords and selects neighbours from the
retrieval history~\cite{wang2024kgp}, GNN-Ret propagates relevance over
keyword graphs with a graph neural network~\cite{li2025gnnret}, HopRAG links
passages through pseudo-queries~\cite{liu2025hoprag}, and CuriousLLM uses
accumulated evidence to generate follow-up queries and decide when to
stop~\cite{yang2025curiousllm}.

EAGR differs from these approaches in three ways.
(i)~\textbf{No materialized graph:} adjacency is induced on demand from
descriptors stored as ordinary VectorDB metadata, so no edge set, graph
index, GNN or pseudo-query index is built.
(ii)~\textbf{Evidence-guided expansion without query rewriting:} the evidence
pooled across explored passages decides which descriptors of the local
frontier to follow next, rather than generating new queries over the whole
corpus or ranking a fixed candidate pool.
(iii)~\textbf{Verified stopping and evidence-based ranking:} traversal stops
only when the LLM's reasoning chain is supported by the collected evidence,
and the retrieved passages are ranked by that chain, rather than by query
similarity alone.
\section{Methodology}
\label{sec:method}
\begin{figure*}
    \centering
    \includegraphics[width=\linewidth]{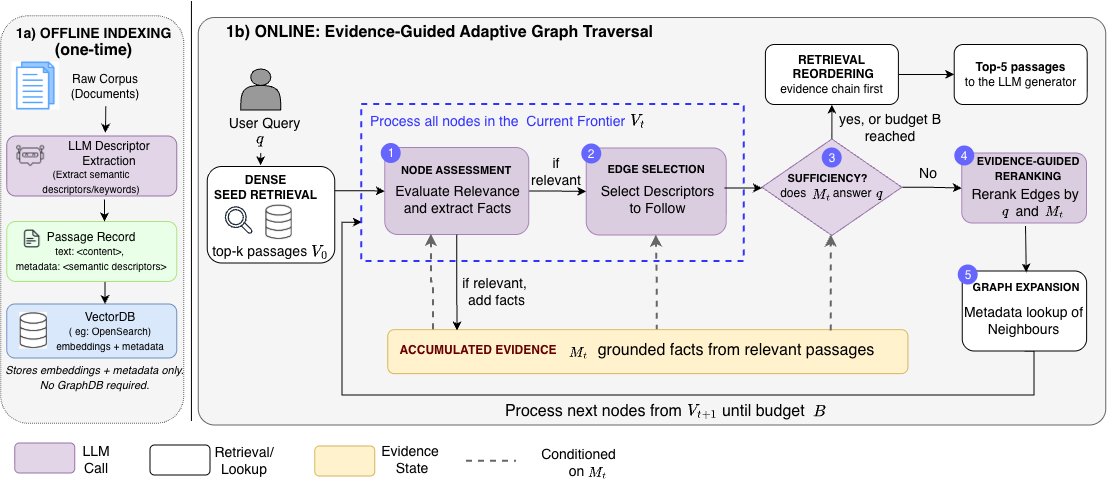}
    \caption{\textbf{Overview of EAGR. (a) Offline indexing:} an LLM extracts semantic descriptors for each passage and stores them as VectorDB metadata; passages sharing a descriptor form an implicit graph. \textbf{(b) Online retrieval:} dense retrieval yields the seed frontier $V_0$. For each frontier node, one LLM call (1) assesses relevance, adding grounded facts to the evidence $M_t$, and (2) selects descriptors to follow. (3) If $M_t$ suffices to answer $q$, traversal stops; otherwise (4) selected descriptors are reranked by $q$ and $M_t$, and (5) neighbours are fetched by metadata lookup, until budget $B$ is reached. Retrieved passages are finally reordered evidence-first.}
    \label{fig:eagr}
\end{figure*}

Figure \ref{fig:eagr} illustrates our overall pipeline, consisting of an offline stage that induces an implicit graph over the corpus (Fig. \ref{fig:eagr}a) and an online stage that traverses this graph via evidence-guided frontier search (Fig. \ref{fig:eagr}b).

\subsection{Problem Formulation}
\label{sec:problem_formulation}

Let $\mathcal{C}=\{p_1,\ldots,p_N\}$ denote a corpus of passages and
$q$ denote an information-seeking query. Standard retrieval in RAG systems
typically formulates retrieval as a mapping from the query to a ranked set of
passages,
\[
\mathcal{R}_{\mathrm{std}}(q,\mathcal{C}) \rightarrow \mathcal{P},
\]
where $\mathcal{P}$ denotes the retrieved passages. Under this formulation,
the query serves as the primary conditioning signal throughout retrieval.
For multi-hop queries, however, the evidence needed to answer $q$ is spread across several passages, and what earlier retrieval uncovers often determines what must be retrieved next. For example, the passage that identifies a film's director is what reveals which biography to retrieve. Query-only retrieval does not model this evolving context.

We therefore treat \textit{adaptive retrieval as a first-class retrieval
problem}, in which each retrieval step is conditioned jointly on the original
query $q$ and the evidence accumulated so far. Let $\mathcal{M}_t$ denote the
evidence accumulated after $t$ retrieval iterations. The passages retrieved
next depend on both:
\[
\mathcal{P}_{t+1} = \mathcal{R}_{\mathrm{adapt}}(q,\mathcal{M}_t,\mathcal{C}),
\]
\[
\mathcal{M}_{t+1} = \mathcal{M}_t \cup \operatorname{Evidence}(\mathcal{P}_{t+1}),
\]
where $\operatorname{Evidence}(\cdot)$ extracts the facts in the newly
retrieved passages that bear on $q$. Retrieval thus alternates between
retrieving passages and updating the evidence state, and
$\mathcal{R}_{\mathrm{adapt}}$ need not search all of $\mathcal{C}$ at every
step: in EAGR it searches only the neighbourhood of the passages explored so
far.

The objective is to acquire sufficient evidence to answer $q$ while
avoiding unnecessary retrieval. Given a retrieval budget $B$, defined as the
maximum number of nodes that may be explored, the process terminates when the
accumulated evidence is sufficient to answer $q$, or when the
exploration budget is exhausted, and returns a relevance-based reordered list of passages. Unlike iterative methods \cite{yao2023react} that issue new model-generated queries at each step, our formulation keeps the
original query $q$ fixed throughout retrieval. Adaptation happens instead
through the evolving evidence state $\mathcal{M}_t$, which determines what
to seek next without altering the original information need.
 
In the following sections, we instantiate this formulation through an
implicit semantic graph abstraction \ref{sec:offline_gc} and an evidence-guided adaptive graph retrieval procedure \ref{sec:eagr}.

\subsection{Offline Stage: Implicit Graph Construction (Figure~\ref{fig:eagr}a)}
\label{sec:offline_gc}

To enable adaptive retrieval beyond the initial set of retrieved passages, 
we introduce an \textit{implicit semantic graph abstraction} over the corpus. 
Rather than explicitly constructing and storing a graph, we attach to each passage $p_i \in \mathcal{C}$ a set of semantic descriptors $K_i = \{k_{i1}, \ldots, k_{im_i}\}$, the entities and key concepts it mentions, and store them as metadata alongside the passage in the VectorDB. Each passage is thus represented as $v_i = (p_i, K_i)$.

Two passages are connected if they share a
descriptor~\cite{wang2024kgp,li2025gnnret}, which defines the implicit graph
$\mathcal{G} = (\mathcal{C}, \mathcal{E})$ with
\[
(p_i,p_j) \in \mathcal{E} \quad \Longleftrightarrow \quad K_i \cap K_j \neq \emptyset .
\]
$\mathcal{G}$ is never materialized: EAGR stores no edges, triples,
hierarchical summaries or auxiliary graph index. Instead, the neighbourhood associated with a selected set of descriptors (K') is retrieved at query time through a metadata lookup in the VectorDB,
\[
\mathcal{N}(K') = \{\, p_i \in \mathcal{C} \mid K_i \cap K' \neq \emptyset \,\},
\]
where descriptors are matched as phrases (Appendix~\ref{app:graph-construction}).
The graph is therefore carried entirely by passage metadata, so graph-based
retrieval needs no separate graph construction, storage or indexing pipeline.
This simplicity is \emph{architectural} rather than computational: EAGR still
issues LLM calls for offline descriptor extraction and online traversal, which
we profile in Section~\ref{sec:results}.

\paragraph{Descriptor Extraction and Adjacency.}
A single LLM call per passage extracts the named entities it mentions (people,
organizations, locations, titles of works, events, products) and a few key
concepts. Descriptors are lowercased, stripped of accents and punctuation, and
kept only if they occur in the passage, which removes hallucinated
descriptors. Dates are not used as descriptors, since they connect many
unrelated passages. Finally, because the same person is often named
differently across passages (e.g., \emph{Clarence Brown} in a film's passage
and \emph{Clarence Leon Brown} in his biography), a corpus-level step links
shorter forms of a person's name to the full name, so that such passages
become neighbours. Appendix~\ref{app:graph-construction} gives the prompt and
the complete procedure.

\subsection{Online Stage: Evidence-Guided Frontier Search (Figure~\ref{fig:overview}b)}
\label{sec:eagr}

Given a query $q$, EAGR first retrieves the top-$k$ passages by dense passage
retrieval (DPR).  Each passage is represented as $v_i=(p_i,K_i)$, with content
$p_i$ and semantic descriptors $K_i$. These seed passages form the initial
frontier $V_0$, and the accumulated evidence is initialized as
$\mathcal{M}_0=\emptyset$. EAGR then proceeds frontier by frontier: it
assesses every node of the current frontier $V_t$, adding the evidence they
contain to $\mathcal{M}_t$, and uses this evidence to decide whether to stop
and, if not, which directions of the implicit graph to explore next. Each
iteration comprises five stages: \textit{Node Assessment and Evidence Accumulation}, \textit{Edge
Selection}, \textit{Sufficiency Check}, \textit{Evidence-Guided Edge
Reranking} and \textit{Graph Expansion}. Node Assessment and Edge Selection
are performed jointly in a single LLM call per node; the prompts are given in
Appendix~\ref{app:prompts}.

\paragraph{Node Assessment and Evidence Accumulation.}
Nodes of the frontier are assessed in order of their similarity to $q$. For
each node $v_i$, EAGR decides whether it is relevant to the query and
extracts the evidence it contributes, conditioned on the query and the
evidence accumulated so far:
\[
(r_i,E_i) = \operatorname{Assess}(q,v_i,\mathcal{M}_t).
\]
A node is relevant ($r_i=1$) if it helps answer any step of the query: it
states information the query asks for, or identifies an entity on the path to
the answer, such as the director of a film named in the query. It need not
contain the final answer. $E_i$ is the set of
facts from $p_i$ that fill a step of the query, each written as a short,
self-contained sentence; facts about the same entities that would not change
the answer are left out. Since the LLM may introduce information that is not
in the passage, a fact is kept only if it is grounded in $p_i$, and facts
already in $\mathcal{M}_t$ are skipped (Appendix~\ref{app:prompts}). The
evidence of relevant nodes is accumulated,
\[
\mathcal{M}_t \leftarrow \mathcal{M}_t \cup E_i \quad \text{if } r_i=1,
\]
so each subsequent node, within the same frontier or a later one, is assessed
against all evidence found so far. Irrelevant nodes contribute neither
evidence nor directions for exploration.

\paragraph{Edge Selection.}
For each relevant node $v_i=(p_i,K_i)$, EAGR considers its unvisited
descriptors as candidate directions for further exploration and selects those
most likely to lead to evidence that is still missing, conditioned on the
query, the current passage and the accumulated evidence:
\[
K_i' = \operatorname{Select}(q,v_i,K_i,\mathcal{M}_t), \qquad K_i' \subseteq K_i .
\]
Descriptors are thus not chosen solely for their association with the current
passage; their utility is judged against what the query still requires given
the evidence gathered so far. Each descriptor in $K_i'$ defines a direction of
traversal, with the passages sharing it forming candidate neighbours in the
implicit graph. By selecting directions rather than exploring all descriptors
in $K_i$, EAGR focuses expansion on connections expected to yield useful
evidence.

Unlike methods that traverse a knowledge graph through predefined entities
and relations, EAGR selects edges using the full content of the current
passage together with the accumulated evidence. This lets it distinguish
directions that are structurally connected but differ substantially in their
usefulness for the query.

\paragraph{Sufficiency Check.}
Once all nodes of the frontier have been assessed, and if they added new
evidence, EAGR checks whether $\mathcal{M}_t$ is sufficient to answer $q$:
\[
s_t = \operatorname{Sufficient}(q,\mathcal{M}_t).
\]
Rather than asking for a bare yes/no judgement, which we found prone to
premature termination, the LLM writes out the chain of steps the query
requires, each as an (\emph{entity}, \emph{attribute}, \emph{value}) triple
with the value taken from $\mathcal{M}_t$, or left empty if the evidence does
not state it. For example, for \emph{``Which team did Joe Buck's father
broadcast for?''}, the chain is (Joe Buck, father, Jack Buck) followed by
(Jack Buck, team, St.\ Louis Cardinals). EAGR accepts $s_t=1$ only if the LLM
judges the evidence sufficient, every value is filled, and every value
appears in $\mathcal{M}_t$, which rejects answers the LLM fills in from its
own knowledge. If $s_t=1$, retrieval terminates; otherwise EAGR continues with
the edges selected in this frontier. Retrieval also terminates when the
number of explored passages reaches the budget $B$.

\paragraph{Evidence-Guided Edge Reranking.}
If the evidence is insufficient, EAGR pools the descriptors selected from relevant nodes on the current frontier into a candidate set:
\[
\mathcal{K}_{\mathrm{cand}} = \bigcup_{v_i \in \mathcal{V}_t} K_i',
\]
where $\mathcal{V}_t \subseteq V_t$ denotes the relevant nodes of the
frontier. It then ranks these candidates by their usefulness for the query,
conditioned on the accumulated evidence:
\[
K^{*} = \operatorname{Rank}\left(q,\mathcal{M}_t,\mathcal{K}_{\mathrm{cand}}\right).
\]
Unlike Edge Selection, which evaluates each node's descriptors independently in its local context, reranking compares candidates proposed by different nodes against the same accumulated evidence. This enables EAGR to prioritize directions that address unresolved evidence gaps and adapt its focus as new evidence is gathered.
\begin{table*}[]
\centering
\small
\resizebox{\textwidth}{!}{
\begin{tabular}{lcccccccc}
\toprule
\multirow{2}{*}{\textbf{Method}} & \multicolumn{2}{c}{\textbf{HotpotQA}} & \multicolumn{2}{c}{\textbf{2WikiMultiHopQA}} & \multicolumn{2}{c}{\textbf{MuSiQue}} & \multicolumn{2}{c}{\textbf{Average}} \\
\cmidrule(lr){2-3} \cmidrule(lr){4-5} \cmidrule(lr){6-7} \cmidrule(lr){8-9}
& \textbf{Str-Acc.} & \textbf{LLM-Acc.} & \textbf{Str-Acc.} & \textbf{LLM-Acc.} & \textbf{Str-Acc.} & \textbf{LLM-Acc.} & \textbf{Str-Acc.} & \textbf{LLM-Acc.} \\
\midrule
\multicolumn{9}{l}{\textit{\textbf{Direct LLM}}} \\
Llama-3.3-70B-Instruct & 41.2 & 49.2 & 43.6 & 43.8 & 17.6 & 20.2 & 34.1 & 37.7 \\
\midrule
\multicolumn{9}{l}{\textit{\textbf{Dense Retrieval}}} \\
Vanilla RAG & 66.4 & 75.5 & 64.2 & 62.1 & 33.7 & 41.9 & 54.8 & 59.8 \\
\midrule
\multicolumn{9}{l}{\textit{\textbf{Graph-aware Retrieval}}} \\
HippoRAG2 & \textbf{68.9} & \textbf{77.9} & 74.9 & 72.7 & 32.9 & 38.7 & 58.9 & 63.1 \\
LinearRAG & 63.8 & 73.6 & \underline{80.1} & \textbf{78.2} & 31.8 & 37.8 & 58.6 & 63.2 \\
MemGraphRAG & 66.6 & 75.5 & 66.7 & 64.6 & 37.0 & 44.2 & 56.8 & 61.4 \\
LogicRAG & 55.7 & 62.4 & 39.3 & 36.7 & 19.9 & 23.6 & 38.3 & 40.9 \\
\midrule
\multicolumn{9}{l}{\textit{\textbf{Agentic / Iterative Retrieval}}} \\
ReAct & 42.1 & 49.3 & 44.1 & 44.5 & 18.0 & 21.9 & 34.7 & 38.6 \\
\midrule
\multicolumn{9}{l}{\textit{\textbf{Adaptive Graph Retrieval (Ours)}}} \\
EAGR + Accum. Evidence & 66.8 & 75.6 & 67.2 & 65.2 & \textbf{43.2} & \textbf{51.7} & \underline{59.1} & \underline{64.2} \\
EAGR + Passage (\sys) & \underline{67.4} & \underline{77.4} & \textbf{80.3} & \underline{78.1} & \underline{38.4} & \underline{45.8} & \textbf{62.0} & \textbf{67.1} \\
\bottomrule
\end{tabular}
}
\caption{End-to-end question answering performance on multi-hop QA benchmarks. We report String-based accuracy (Str-Acc.) and LLM judge accuracy (LLM-Acc.). The \textbf{best} result in each column is shown in bold, and the \underline{second-best} is underlined.}
\label{tab:main_results}
\end{table*}
\paragraph{Graph Expansion.}
Given the ranked descriptors $K^{*}$, EAGR materializes the corresponding
neighbourhood $\mathcal{N}(K^{*})$ through a metadata lookup in the VectorDB
(\S\ref{sec:offline_gc}), retrieving up to $N_{\max}$ passages per descriptor.
Already visited passages are excluded, and the remaining passages form the
next frontier $V_{t+1}$, ordered by the rank of the descriptor that reached
them; since the budget may run out partway through a frontier, higher-ranked
directions are explored first. Implementation details of candidate edges,
such as following other spellings of a selected name, are given in
Appendix~\ref{app:prompts}. The process then iterates, so that the evidence
accumulated at each step shapes the directions explored next, without an
explicit graph ever being constructed or stored.

\paragraph{Termination and output.}
Traversal stops when the Sufficiency Check succeeds or when the number of
explored passages reaches the budget $B$. The number of passages explored
therefore varies across queries: EAGR retrieves only as much additional
evidence as each query needs, within a fixed upper bound on retrieval cost.
The explored passages are then ranked as described in
\S\ref{sec:retrieval_reordering}.

All semantic operations in EAGR use three prompts to the same LLM: a joint
prompt for Node Assessment and Edge Selection (one call per node), a
Sufficiency Check prompt (at most one call per frontier) and an Edge
Reranking prompt (at most one call per frontier). The prompts are given in
Appendix~\ref{app:prompts}, a step-by-step retrieval trace in
Appendix~\ref{app:retrieval-trace}, and the complete procedure in
Algorithm~\ref{appendix:alg-eagr}.

\subsection{Retrieval Reordering}
\label{sec:retrieval_reordering}

After traversal terminates, EAGR orders the explored passages using the
evidence collected along the way, in three groups:
(i)~relevant passages whose extracted facts mention the entities or values in
the reasoning chain of the last Sufficiency Check, i.e., the evidence for the
answer;
(ii)~other relevant passages that contributed facts; and
(iii)~all remaining passages, with those judged irrelevant last.
Within each group, passages keep their order from the traversal.

This ordering matters most for multi-hop queries, where passages reached
after one or more hops (e.g., the biography of a film's director) are needed
to establish the answer but may share little vocabulary with the query.
Ranked by query similarity alone, such passages would often fall outside the
top-ranked results; ranking by the evidence chain instead places them first,
while passages judged irrelevant are moved to the end.

\section{Experiments}
\label{sec:experimentals}

In this section, we aim to answer the following research questions: \textit{Q1 (Generation Accuracy):} How does EAGR compare with state-of-the-art retrieval-augmented generation and GraphRAG methods in terms of end-to-end answer generation performance? \textit{Q2 (Retrieval Effectiveness):}  Does \sys{} retrieve the supporting passages more
effectively, particularly for multi-hop information needs? \textit{Q3 (Ablation Study):} What is the contribution of each component of EAGR contribute to the overall retrieval and generation performance?

\subsection{Experimental Setup}
\label{sec:experimental_setup}
\begin{table*}[ht]
\centering
\scriptsize
\resizebox{\textwidth}{!}{
\begin{tabular}{lcccccccc}
\toprule
\multirow{2}{*}{\textbf{Method}} & \multicolumn{2}{c}{\textbf{HotpotQA}} & \multicolumn{2}{c}{\textbf{2WikiMultiHopQA}} & \multicolumn{2}{c}{\textbf{MuSiQue}} & \multicolumn{2}{c}{\textbf{Average}} \\
\cmidrule(lr){2-3} \cmidrule(lr){4-5} \cmidrule(lr){6-7} \cmidrule(lr){8-9}
& \textbf{Str-Acc.} & \textbf{LLM-Acc.} & \textbf{Str-Acc.} & \textbf{LLM-Acc.} & \textbf{Str-Acc.} & \textbf{LLM-Acc.} & \textbf{Str-Acc.} & \textbf{LLM-Acc.} \\
\midrule
\textbf{EAGR (Full)} & 67.4 & 77.4 & \underline{80.3} & \underline{78.1} & \underline{38.4} & \underline{45.8} & \underline{62.0} & \textbf{67.1} \\
\midrule
\multicolumn{9}{l}{\textit{\textbf{Evidence Assessment}}} \\
\quad w/o Relevance Scoring & \textbf{67.6} & \underline{77.5} & 79.0 & 76.4 & \textbf{39.6} & \textbf{46.1} & \textbf{62.1} & \underline{66.7} \\
\quad w/o Accumulated Evidence & 67.1 & 76.1 & 74.5 & 71.8 & 34.7 & 40.4 & 58.8 & 62.8 \\
\midrule
\multicolumn{9}{l}{\textit{\textbf{Graph Navigation}}} \\
\quad w/o Edge Selection & 40.9 & 48.3 & 44.4 & 44.6 & 32.5 & 38.5 & 39.3 & 43.8 \\
\quad w/o Edge Reranking & \underline{67.5} & \textbf{78.1} & \textbf{80.4} & \textbf{78.2} & 36.6 & 42.6 & 61.5 & 66.3 \\
\midrule
\multicolumn{9}{l}{\textit{\textbf{Retrieval Policy}}} \\
\quad w/o Retrieval Reordering & 65.8 & 74.6 & 73.6 & 70.2 & 36.8 & 44.8 & 58.7 & 63.2 \\
\bottomrule
\end{tabular}
}
\caption{Ablation study of EAGR components across HotpotQA, 2WikiMultiHopQA, and MuSiQue benchmarks. The \textbf{best} result in each column is shown in bold, and the \underline{second-best} is underlined.}
\label{tab:ablation}
\end{table*}

\begin{table*}[t]
\centering
\scriptsize
\renewcommand{\arraystretch}{1.15}
\begin{tabularx}{\textwidth}{l p{7 cm} p{4.2cm} c}
\toprule
\multicolumn{4}{l}{\textbf{Question:} \textit{Who is the owner of the record label of the performer of What Kind of Love?} \quad \textbf{Gold:} \textit{Warner Music Group}} \\
\midrule
\textbf{Method} & \textbf{Reasoning / Retrieval Breakdown} & \textbf{Generated Answer} & \textbf{Judge} \\
\midrule
\textbf{Direct LLM} & Parametric hallucination; assumes Sony Music without retrieval. & \textit{Sony Music Entertainment} & \textcolor{red}{\ding{55}} \\
\textbf{Vanilla RAG} & Dense search retrieves song info (Rodney Crowell) but misses label owner. & \textit{The information is not provided.} & \textcolor{red}{\ding{55}} \\
\textbf{HippoRAG2} & PPR graph activation disperses across artist albums, missing the owner. & \textit{The information is not provided...} & \textcolor{red}{\ding{55}} \\
\textbf{MemGraphRAG} & Memory graph clusters song details but fails to jump to parent owner entity. & \textit{The information is not provided...} & \textcolor{red}{\ding{55}} \\
\textbf{LogicRAG} & Logical decomposition resolves performer but fails to retrieve 3rd-hop owner. & \textit{The information is not provided.} & \textcolor{red}{\ding{55}} \\
\textbf{LinearRAG} & Linear graph misses multi-hop connectivity between artist and holding entity. & \textit{The information is not provided.} & \textcolor{red}{\ding{55}} \\
\textbf{ReAct} & Agent repeats sub-query formulation, entering an empty retrieval loop. & \textit{[Empty]} & \textcolor{red}{\ding{55}} \\
\midrule
\rowcolor{gray!12}
\textbf{EAGR (Ours)} & \textbf{Hop 1:} \textit{What Kind of Love} $\rightarrow$ \textbf{Hop 2:} \textit{Rodney Crowell} $\rightarrow$ \textbf{Hop 3:} \textit{Warner Bros.\ Records} $\rightarrow$ \textit{Warner Music Group} (Terminates on sufficiency). & \textbf{\textit{Warner Music Group}} & \textcolor{green!60!black}{\textbf{\checkmark}} \\
\bottomrule
\end{tabularx}
\caption{Qualitative comparison on MuSiQue (\texttt{3hop1\_\_198276\_709625\_84283}). EAGR recovers the full 3-hop reasoning chain, whereas baselines fail due to missing context hops, search diffusion, or loop breakdown.}
\label{tab:qualitative_case_study}
\end{table*}
\paragraph{Datasets.}
We evaluate \sys{} on three widely used multi-hop QA benchmarks:
HotpotQA, 2WikiMultiHopQA (2Wiki), and MuSiQue
~\cite{yang2018hotpotqa,ho2020twowiki,trivedi2022musique}. Following prior
work~\cite{gutierrez2024hipporag,gutierrez2025hipporag2}, we select 1,000 validation questions from
each dataset.

\paragraph{Baselines.}
We compare against direct generation without retrieval, vanilla dense RAG,
the graph-based methods HippoRAG2~\cite{gutierrez2025hipporag2},
LinearRAG~\cite{zhuang2026linearrag}, MemGraphRAG~\cite{memgraphrag} and
LogicRAG~\cite{logicrag}, and the iterative methods
ReAct~\cite{yao2023react}. ReAct use the same retriever, LLM and passage budget as \sys{}.

\paragraph{Evaluation Metrics.}
We evaluate EAGR from both retrieval and downstream QA perspectives. For retrieval, we report Recall@5, the fraction of gold supporting passages among the top five retrieved, computed on each method's own ranking. For QA, every method's top five passages are given to the same LLM generator; String Accuracy (Str-Acc.) checks whether the normalized gold answer appears in
the response, and LLM Accuracy (LLM-Acc.) uses an LLM
judge~\cite{memgraphrag}.

\paragraph{Implementation Details.}
Dense retrieval uses \textsc{NV-Embed-v2}~\cite{lee2025nvembed} for all
methods. Descriptors are extracted with
\textsc{Llama-3.3-70B-Instruct}~\cite{meta2024llama33}, and \sys{}'s retrieval
prompts use \textsc{Qwen3-8B}~\cite{qwen3technicalreport} with thinking
disabled. \sys{} starts from $k=4$ seed passages, with a budget of $B=30$
passages. Answers are generated by \textsc{Llama-3.3-70B-Instruct} and judged
by \textsc{DeepSeek-V3.2}~\cite{liu2025deepseek}. All LLMs use temperature 0. We also report a variant whose reader receives the accumulated evidence $M_t$ instead of passages.

\section{Results and Analysis}
\label{sec:results}

\paragraph{Generation accuracy (Q1).}
Table~\ref{tab:main_results} shows that \sys{} achieves the highest average
accuracy, 67.1 LLM-Acc., 3.9 points above the strongest baseline (LinearRAG,
63.2) and 7.3 above vanilla RAG. Its main strength is consistency: the
strongest graph-based baselines each excel on some benchmarks but fall below
vanilla RAG on others (HippoRAG2 and LinearRAG on MuSiQue, LinearRAG also on
HotpotQA), whereas \sys{} improves over vanilla RAG on all three. It matches
the best baseline on HotpotQA (77.4 vs.\ 77.9) and 2Wiki (78.1 vs.\ 78.2) and
gives the best result on MuSiQue (45.8 vs.\ 44.2), the benchmark with the
longest reasoning chains. Giving the genertor the accumulated evidence instead of
passages raises MuSiQue accuracy further to 51.7, suggesting that the evidence
collected during traversal is a useful condensed context for long chains.

\paragraph{Retrieval (Q2).}
Table~\ref{tab:retrieval_recall5_single} shows the largest retrieval gain on 2Wiki, where \sys{}
reaches 95.5 Recall@5 against 91.0 for HippoRAG2 and 76.7 for dense
retrieval; its questions mostly follow explicit entity chains (e.g., a film's
director, then the director's birthplace), which descriptor edges follow
directly. On HotpotQA, \sys{} is on par with HippoRAG2 (95.4 vs.\ 95.9). On
MuSiQue, its recall equals dense retrieval and trails HippoRAG2 (71.5 vs.\
76.2), yet its answer accuracy is 7.1 points higher, indicating that recall of
all gold passages alone does not fully determine downstream accuracy. Figure~\ref{fig:early_stopping} shows how early stopping criteria in \sys{} makes a significant number of queries stop at less than 5 retrieved chunks, this filtering out noisy context and providing better QA accuracy. 

\paragraph{Adaptive retrieval.}
Because traversal stops once the accumulated evidence is verified to answer
the query, \sys{} spends retrieval effort where it is needed
(Figure~\ref{fig:early_stopping}). On HotpotQA and 2Wiki, about 81\% of
queries stop within 10 passages, and on HotpotQA 71\% need at most 5, i.e.,
the four seed passages and at most one more; only about 4\% reach the budget.
MuSiQue, whose questions require up to four hops, shows a different profile:
about 46\% of queries stop within 10 passages, while 31\% explore the full
budget. The traversal thus scales with question difficulty rather than
retrieving a fixed number of passages per query, which keeps the average
cost low on benchmarks dominated by short reasoning chains.

\subsection{Ablation Study}
\label{sec:ablation}
Each ablation removes one component while keeping the rest of the pipeline
unchanged (Table~\ref{tab:ablation}). Two components matter consistently.
Removing accumulated evidence, so that assessment, edge selection and
reranking see only the query and the current passage and traversal never
stops early, lowers average accuracy by 4.3 points, and by 6.3 and 5.4 on 2Wiki
and MuSiQue: conditioning on evidence is the main source of \sys{}'s gains.
Removing retrieval reordering costs 3.9 points (7.9 on 2Wiki), showing that
passages found through multi-hop expansion need to be ranked by the evidence
chain to reach the reader. Edge reranking has a small average effect (0.8) but
matters on MuSiQue (3.2), where longer chains make the order of exploration
more consequential. Relevance assessment has the smallest effect: it helps on
2Wiki (1.7) and is neutral on HotpotQA and MuSiQue.
Edge selection is the most critical component: following every descriptor of
a relevant passage instead of the selected ones lowers average accuracy by
23.3 points, below even vanilla RAG. Without selection, the budget is spent
within the first expansion on passages that share incidental descriptors
with the seeds, so the passages of later hops are rarely reached, and the
many loosely related passages explored compete with the actual evidence for
the top-ranked positions.


\subsection{Qualitative Analysis}
\label{sec:qualitative_analysis}
To examine how evidence guides traversal, we analyze a 3-hop MuSiQue query where all baselines fail (Table~\ref{tab:qualitative_case_study}, Appendix~\ref{sec:appendix_qualitative}). While dense, graph, and agentic baselines fail to bridge intermediate entities (\textit{Rodney Crowell} $\rightarrow$ \textit{Warner Bros.\ Records} $\rightarrow$ \textit{Warner Music Group}) due to localization bias, hallucination, or looping, EAGR's Node Assessment iteratively accumulates evidence across hops and successfully recovers the complete reasoning chain.

\subsection{Human Evaluation}
\label{sec:exp_human_eval}
To validate the LLM judge, three annotators independently labelled 150
sampled predictions across benchmarks and methods. The judge agreed with the
human majority on 94.0\% of them (Fleiss' $\kappa = 0.854$;
Appendix~\ref{sec:human_eval}).

\section{Conclusion}
\label{sec:concl}
\eat{
We presented EAGR, an evidence-guided approach to adaptive retrieval that
moves beyond query-centric retrieval by allowing accumulated evidence to
guide subsequent retrieval while keeping the original query fixed. EAGR
induces an implicit semantic graph from passage descriptors and navigates it
through Node Assessment, Edge Selection, Evidence-Guided Edge Reranking, and
Graph Expansion, without requiring explicit graph construction. Experiments
across multi-hop and graph-centric benchmarks demonstrate its effectiveness,
with particularly strong gains on multi-hop question answering. Overall, our
results demonstrate that accumulated evidence provides an effective
conditioning signal for adaptive retrieval, enabling systems to better follow
the information needed to answer complex queries.}

We presented EAGR, an adaptive retrieval approach in which accumulated
evidence, rather than a rewritten query, determines which directions to
explore next. EAGR traverses an implicit graph induced from descriptors stored
as ordinary VectorDB metadata, selecting and ranking edges by the evidence
collected and stopping once a verified reasoning chain answers the query, all
without materializing a graph or rewriting the query. Across HotpotQA,
2WikiMultiHopQA and MuSiQue, it achieves the highest average answer accuracy,
3.9 points above the strongest graph-based baseline, and is the only
graph-based method that improves over dense retrieval on all three
benchmarks. Ablations show that conditioning on accumulated evidence drives
these gains, supporting a view of multi-hop retrieval as an adaptive search process in which accumulated evidence and not the original query alone, determines how the retrieval frontier evolves. 

\section{Limitations}
All components of EAGR are currently used off-the-shelf without
task-specific training. Future work could explore component-specific
fine-tuning and stronger foundation models to further improve both retrieval
quality and practical viability.

\section{Ethics Statement}
We used LLM-based tools to assist with code development and to improve the
clarity of the writing. All research ideas, the method, the experimental
design, and the analysis and interpretation of results are the authors' own,
and all LLM-assisted text was verified by the authors for correctness.

\paragraph{Reproducibility.}
EAGR builds only on publicly available datasets and open-weight models. Our
experiments use the standard HotpotQA, 2WikiMultiHopQA, MuSiQue, and
GraphRAG-Bench evaluation sets with the model configuration reported in
Section~\ref{sec:experimental_setup}, and all prompts used by the offline and
online stages are provided verbatim in Appendices~\ref{app:graph-construction}
and~\ref{app:prompts}. Upon acceptance, we will release the code and the
generated passage metadata for all datasets used to enable exact reproduction
of the implicit graph and the retrieval procedure.

\bibliography{custom}

\appendix

\section{Appendix}
\label{sec:appendix}
\subsection{Evidence-Guided Adaptive Graph Retrieval (EAGR) Algorithm}
Algorithm~\ref{appendix:alg-eagr} provides the complete procedural description of
EAGR. The algorithm consists of an offline semantic graph abstraction stage
and an online evidence-guided retrieval stage. During the offline stage, each
passage is associated with a set of semantic descriptors $K_i$, which are
stored as metadata in the VectorDB. During inference, EAGR first retrieves
$k$ seed passages using DPR and initializes the retrieval queue and accumulated
evidence with these seeds and an empty evidence context, respectively.

The online retrieval procedure follows the four stages described in
Section~\ref{sec:eagr}: Node Assessment, Edge Selection,
Evidence-Guided Edge Reranking, and Graph Expansion. Relevant but insufficient
nodes contribute evidence and selected semantic descriptors to the traversal,
while irrelevant nodes are excluded from further expansion. Candidate
descriptors proposed across the explored nodes are subsequently reranked using
the query and accumulated evidence, after which at most $N_{\max}$ neighboring
passages are retrieved through metadata lookup. Already explored nodes are
excluded from subsequent expansion.

The procedure terminates when a relevant node is assessed as sufficient or
when the exploration budget $B$ is exhausted. Finally, the retrieved passages
are reordered using the relevance decisions made during traversal, while
preserving their relative retrieval order.

\section{Multi-Hop Retrieval Performance}
\label{sec:appendix_retrieval}

Table~\ref{tab:retrieval_recall5_single} reports the retrieval quality of EAGR compared to dense and graph-aware baselines on multi-hop benchmarks using Proportion Recall at top-5 retrieved passages (\textbf{Recall@5}). 

Operating strictly over implicit metadata stored within a standard VectorDB, EAGR achieves competitive retrieval quality across all benchmarks ($94.5\%$ on HotpotQA, $90.4\%$ on 2WikiMultiHopQA, and $72.6\%$ on MuSiQue), demonstrating that iterative evidence-guided frontier expansion successfully gathers multi-hop supporting context without requiring explicit GraphDB materialization.

\begin{table}[t]
\centering
\small
\resizebox{\columnwidth}{!}{
\begin{tabular}{l ccc}
\toprule
\textbf{Method} & \textbf{HotpotQA} & \textbf{2WikiMHQA} & \textbf{MuSiQue} \\
\midrule
\multicolumn{4}{l}{\textit{\textbf{Dense Retrieval}}} \\
Vanilla RAG (Top-5) & 94.03 & 76.67 & 71.42 \\
\midrule
\multicolumn{4}{l}{\textit{\textbf{Graph-Aware Retrieval}}} \\
LogicRAG            & 68.40 & 45.10 & 20.90 \\
HippoRAG2           & 95.90 & 91.03 & 76.17 \\
LinearRAG           & 66.85 & 80.70 & 65.90 \\
MemGraphRAG         & 92.30 & 80.20 & 71.40 \\
\midrule
\multicolumn{4}{l}{\textit{\textbf{Adaptive Graph Retrieval}}} \\
\rowcolor{gray!12}
\textbf{EAGR (Ours)} & 95.40 & 95.53 & 71.53 \\
\bottomrule
\end{tabular}
}
\caption{Retrieval performance on multi-hop QA benchmarks measured by Proportion Recall at Top-5 (\textbf{Recall@5}).}
\label{tab:retrieval_recall5_single}
\end{table}

\begin{figure}[t]
\centering
\includegraphics[width=\linewidth]{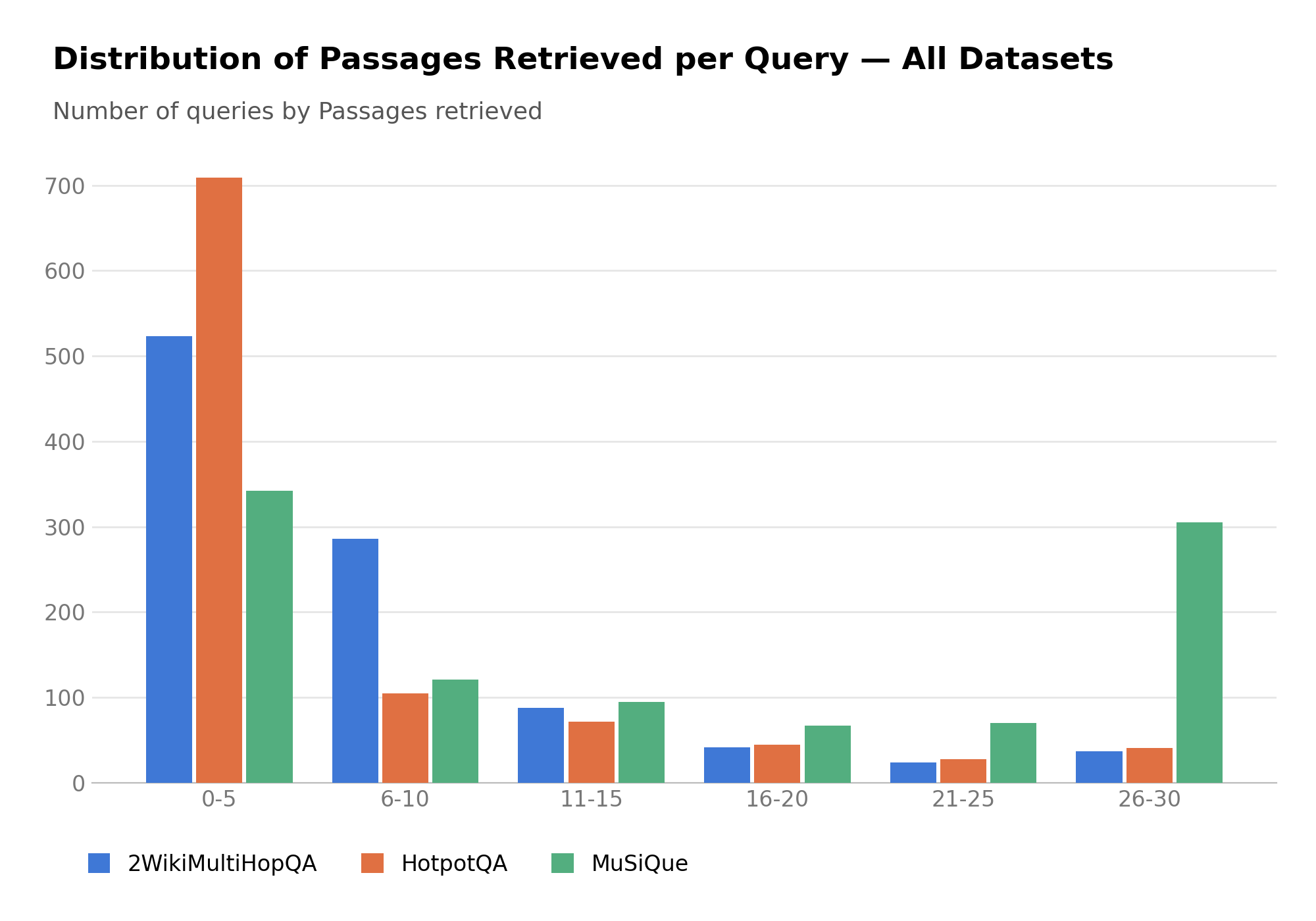}
\caption{Number of passages retrieved per query by \sys{} (budget $B=30$),
per benchmark. Traversal stops as soon as the accumulated evidence is
verified to answer the query.}
\label{fig:early_stopping}
\end{figure}

\section{Qualitative Case Study}
\label{sec:appendix_qualitative}

Table~\ref{tab:qualitative_case_study} presents a detailed qualitative case study on a representative 3-hop question from MuSiQue (\texttt{3hop1\_\_198276\_709625\_84283}), comparing EAGR against all baselines.

\subsection{Case Study: 4-Hop Multi-Entity Disambiguation}
\label{sec:appendix_case_4hop}

Table~\ref{tab:qualitative_case_4hop} presents a challenging 4-hop question from MuSiQue (\texttt{4hop3\_\_771862\_\_466199\_\_695123\_\_72134}) involving multiple overlapping entities (multiple towns named \textit{Springfield} across different states).

\begin{table*}[t]
\centering
\small
\resizebox{\textwidth}{!}{
\begin{tabularx}{\textwidth}{l p{7cm} p{4cm} c}
\toprule
\multicolumn{4}{l}{\textbf{Question:} \textit{When did the town WIZE is licensed in become capitol of the state where Jackson Township is located?}} \\
\multicolumn{4}{l}{\textbf{Ground Truth:} \textit{1839}} \\
\multicolumn{4}{l}{\textbf{4-Hop Chain:} \textit{WIZE} $\rightarrow$ \textit{Springfield} $\rightarrow$ \textit{Illinois} (via \textit{Jackson Township}) $\rightarrow$ \textit{Springfield, Illinois} $\rightarrow$ \textit{1839}} \\
\midrule
\textbf{Method} & \textbf{Retrieval Behavior \& Failure Mode} & \textbf{Generated Output} & \textbf{Judge} \\
\midrule
\textbf{Direct LLM} & Parametric memory fails on 4-hop compositional constraints; returns empty. & \textit{[Empty]} & \textcolor{red}{\ding{55}} \\
\textbf{Vanilla RAG} & Dense retrieval pulls conflicting Springfield passages across states; fails to synthesize. & \textit{[Empty]} & \textcolor{red}{\ding{55}} \\
\textbf{HippoRAG2} & Graph PPR spreads activation to distractor node (\textit{Springfield, Ohio}), concluding false premise. & \textit{``Springfield, Ohio has never been the capital of Indiana.''} & \textcolor{red}{\ding{55}} \\
\textbf{MemGraphRAG} & Entity clustering conflates Ohio and Illinois nodes, generating a contradictory claim. & \textit{``Springfield, Ohio, has never been the capital of Ohio.''} & \textcolor{red}{\ding{55}} \\
\textbf{LinearRAG} & Linear graph misses parallel branch binding \textit{Jackson Township} to \textit{Illinois}. & \textit{``Never''} & \textcolor{red}{\ding{55}} \\
\textbf{LogicRAG} & Sub-question decomposition fails to resolve the joint state constraint. & \textit{``There is no information to answer the question.''} & \textcolor{red}{\ding{55}} \\
\textbf{ReAct} & Agent converges on the state formation year of Indiana (1816) instead of capital relocation. & \textit{``1816''} & \textcolor{red}{\ding{55}} \\
\midrule
\rowcolor{gray!12}
\textbf{EAGR (Ours)} & 
\textbf{Hop 1:} \textit{WIZE} $\rightarrow$ town of \textit{Springfield}. \newline
\textbf{Hop 2:} \textit{Jackson Township} $\rightarrow$ state of \textit{Illinois}. \newline
\textbf{Hop 3:} Evidence combines (\textit{Springfield} + \textit{Illinois}) $\rightarrow$ prunes Ohio/Indiana distractors. \newline
\textbf{Hop 4:} Expands to Springfield history $\rightarrow$ retrieves capital relocation in \textit{1839} (Sufficiency reached). & 
\textbf{\textit{``1839''}} & 
\textcolor{green!60!black}{\textbf{\checkmark}} \\
\bottomrule
\end{tabularx}
}
\caption{Qualitative 4-hop case study on MuSiQue (\texttt{4hop3\_\_771862\_\_466199\_\_695123\_\_72134}). Existing graph methods get trapped in spurious entity paths (*Springfield, Ohio / Indiana*), whereas EAGR's accumulated evidence enables entity disambiguation across parallel reasoning branches.}
\label{tab:qualitative_case_4hop}
\end{table*}

\section{Human Evaluation and LLM Judge Validation}
\label{sec:human_eval}
To rigorously validate automated LLM-as-a-judge evaluation, we conducted a blind human evaluation on a stratified sample of 150 question--answer predictions across HotpotQA, 2WikiMultiHopQA, and MuSiQue (50 questions per benchmark), spanning five retrieval methods (10 samples per method per dataset). Three independent human annotators evaluated all 150 instances using a dedicated Label Studio interface, judging model generations against reference gold answers and supporting context.

\begin{figure}[t] \centering \includegraphics[width=\linewidth]{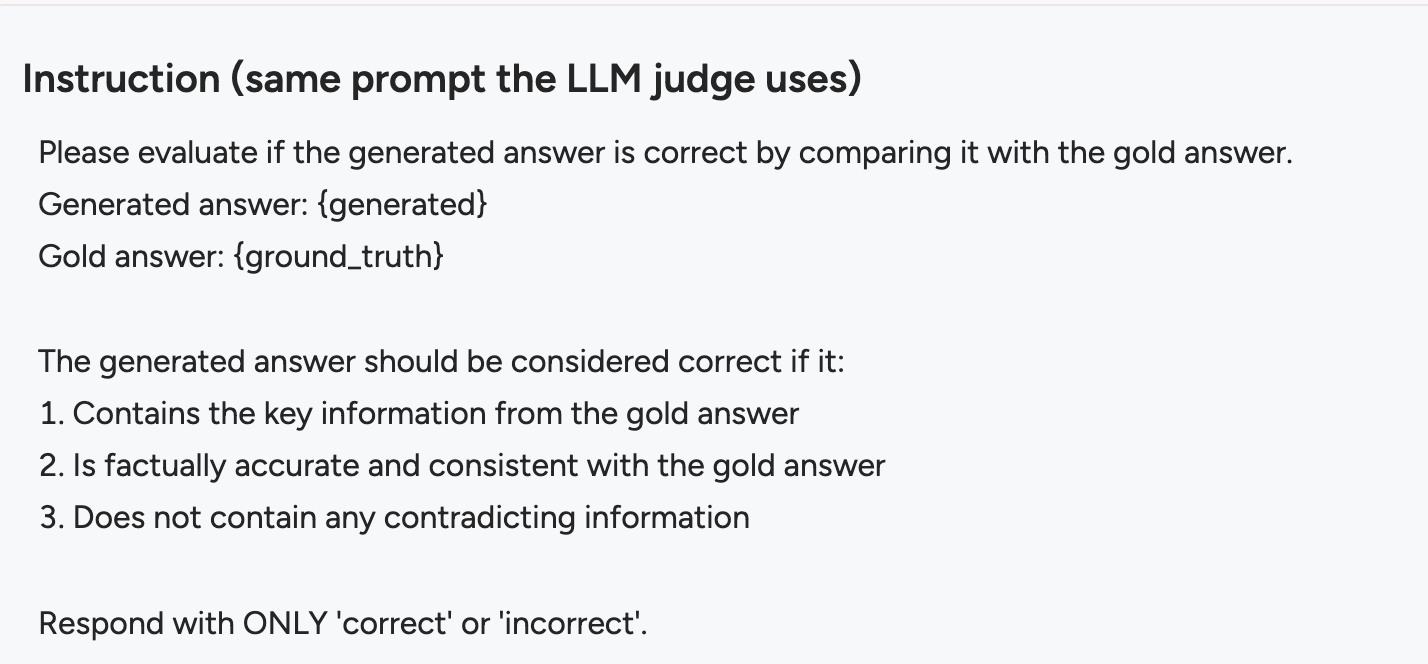} \caption{Instructions provided to human annotators. The same criteria were used for both human evaluation and the LLM judge.} \label{fig:human-eval-prompt} \end{figure}

As detailed in Table~\ref{tab:human_eval}, the three human annotators demonstrated high inter-annotator agreement with a Fleiss' $\kappa$ of $0.844$ (pairwise Cohen's $\kappa$ ranging between $0.822$ and $0.878$). When comparing the automated LLM judge (\texttt{DeepSeek-V3.2}) against the human majority consensus, the judge achieved an overall agreement of $\mathbf{94.0\%}$ (141/150 tasks), with joint Fleiss' $\kappa = 0.854$. Judge agreement remained high across individual benchmarks ($96.0\%$ on both 2WikiMultiHopQA and MuSiQue; $90.0\%$ on HotpotQA) and across retrieval methods ($96.7\%$ on EAGR, HippoRAG2, and LinearRAG). Furthermore, the overall proportion of answers judged correct by the automated judge ($57.3\%$) closely matches human majority consensus ($58.0\%$). Qualitative audit of the 9 dissenting cases indicates that discrepancies stem from borderline partial matches (e.g., shortened names or omitted qualifiers) rather than categorical errors, confirming that the LLM judge reliably mirrors human judgment. The exact instructions provided to human annotators are shown in Figure~\ref{fig:human-eval-prompt}.

\begin{table*}[t]
\centering
\small
\resizebox{\textwidth}{!}{
\begin{tabular}{llr | llr | llr}
\toprule
\multicolumn{3}{c|}{\textbf{(a) Inter-Annotator Agreement ($n=150$)}} & 
\multicolumn{3}{c|}{\textbf{(b) Judge vs.\ Majority by Dataset \& Method}} & 
\multicolumn{3}{c}{\textbf{(c) \% Marked Correct on Sample ($n=150$)}} \\
\midrule
\textbf{Pair / Metric} & \textbf{Agreement} & \textbf{Cohen's $\kappa$} & 
\textbf{Data / Method Slice} & \textbf{Agreement} & \textbf{Count} & 
\textbf{Evaluator} & \textbf{\% Correct} & \textbf{Count} \\
\midrule
Annotator 1 -- Annotator 2 & 92.0\% & 0.833 & HotpotQA & 90.0\% & 45/50 & Annotator 1 & 63.3\% & 95/150 \\
Annotator 1 -- Annotator 3 & 91.3\% & 0.822 & 2WikiMultiHopQA & 96.0\% & 48/50 & Annotator 2 & 58.0\% & 87/150 \\
Annotator 2 -- Annotator 3 & 94.0\% & 0.878 & MuSiQue & 96.0\% & 48/50 & Annotator 3 & 54.7\% & 82/150 \\
\cmidrule{1-3} \cmidrule{4-6} \cmidrule{7-9}
Fleiss' $\kappa$ (3 Annotators) & \multicolumn{2}{c|}{\textbf{0.844}} & Vanilla RAG & 86.7\% & 26/30 & \textbf{Human Majority} & \textbf{58.0\%} & \textbf{87/150} \\
Joint Fleiss' $\kappa$ (+ Judge) & \multicolumn{2}{c|}{\textbf{0.854}} & HippoRAG2 & 96.7\% & 29/30 & \textbf{LLM Judge} & \textbf{57.3\%} & \textbf{86/150} \\
Annotator 1 -- Judge & 90.0\% & 0.792 & LinearRAG & 96.7\% & 29/30 & 3-Human Unanimous & 88.7\% & 133/150 \\
Annotator 2 -- Judge & 94.0\% & 0.877 & MemGraphRAG & 93.3\% & 28/30 & All-4 Unanimous & 87.3\% & 131/150 \\
Annotator 3 -- Judge & 96.0\% & 0.919 & \textbf{EAGR+Passage} & \textbf{96.7\%} & \textbf{29/30} & Disagreements & \phantom{0}6.0\% & \phantom{0}9/150 \\
\midrule
\multicolumn{9}{c}{\textbf{Overall Automated Judge Alignment with Human Majority: \ 94.0\% (141/150) \quad $\vert$ \quad Joint Fleiss' $\kappa = 0.854$}} \\
\bottomrule
\end{tabular}
}
\caption{Human evaluation study ($n=150$) validating automated LLM judging against three independent human annotators across benchmarks and retrieval methods. Annotators achieve high inter-annotator agreement (Fleiss' $\kappa = 0.844$), and the automated judge aligns with human majority consensus on 94.0\% of tasks ($\kappa = 0.854$) with calibrated accuracy ($57.3\%$ vs.\ $58.0\%$).}
\label{tab:human_eval}
\end{table*}

\label{appendix:alg-eagr}
\begin{algorithm*}[t]
\caption{Evidence-Guided Adaptive Graph Retrieval (EAGR)}
\label{appendix:alg-eagr}
\begin{algorithmic}[1]
\Require Query $q$, seed size $k$, budget $B$, neighbour limit $N_{\max}$,
         VectorDB $\mathcal{D}$ storing each passage $p_i$ with descriptors $K_i$
\Ensure Ranked list of retrieved passages
\Statex \textbf{Offline: implicit graph construction} (Appendix~\ref{app:graph-construction})
\For{each passage $p_i \in \mathcal{C}$}
    \State Extract, ground and normalize descriptors $K_i$; store $(p_i, K_i)$ in $\mathcal{D}$
\EndFor
\Statex \textbf{Online: evidence-guided frontier search}
\State $V \gets \textsc{DPR}(q, k)$ \Comment{seed frontier $V_0$}
\State $R \gets [\,]$;\; $\mathcal{M} \gets \emptyset$;\; $U \gets \emptyset$
       \Comment{explored passages, evidence, expanded descriptors}
\While{$V \neq \emptyset$ \textbf{and} $|R| < B$}
    \State $\mathcal{K}_{\mathrm{cand}} \gets \emptyset$;\; $\textit{new} \gets \textbf{false}$
    \For{each $v_i = (p_i, K_i) \in V$ in order of similarity to $q$}
        \If{$|R| \geq B$} \textbf{break} \EndIf
        \State Append $v_i$ to $R$
        \State $(r_i, E_i, K_i') \gets \textsc{AssessAndSelect}(q, v_i, K_i \setminus U, \mathcal{M})$
               \Comment{one LLM call}
        \If{$r_i = 1$}
            \State $E_i \gets \textsc{Ground}(E_i, p_i) \setminus \mathcal{M}$
                   \Comment{drop ungrounded and repeated facts}
            \If{$E_i \neq \emptyset$}
                \State $\mathcal{M} \gets \mathcal{M} \cup E_i$;\; $\textit{new} \gets \textbf{true}$
            \EndIf
            \State $\mathcal{K}_{\mathrm{cand}} \gets \mathcal{K}_{\mathrm{cand}} \cup K_i'$
        \EndIf
    \EndFor
    \If{$\textit{new}$ \textbf{and} $\textsc{Sufficient}(q, \mathcal{M})$}
        \State \textbf{break} \Comment{verified reasoning chain; \S\ref{sec:eagr}}
    \EndIf
    \State $K^{*} \gets \textsc{Rank}(q, \mathcal{M}, \mathcal{K}_{\mathrm{cand}})$;\; $U \gets U \cup K^{*}$
    \State $V \gets \textsc{Neighbours}(K^{*}, N_{\max}, \mathcal{D}) \setminus R$
           \Comment{next frontier, in rank order}
\EndWhile
\State \Return $\textsc{Reorder}(R, \mathcal{M})$ \Comment{\S\ref{sec:retrieval_reordering}}
\end{algorithmic}
\end{algorithm*}
\tcbset{eagrsmall/.style={listing options={basicstyle=\ttfamily\scriptsize,
  breaklines=true, columns=fullflexible,
  literate={–}{{--}}1 {é}{{\'e}}1}}}
  
\definecolor{eagrtraceblue}{RGB}{246,250,252}
\definecolor{eagrtracegray}{RGB}{249,249,249}
\definecolor{eagrtracepeach}{RGB}{253,248,243}
\definecolor{eagrtracegreen}{RGB}{245,251,246}
\definecolor{eagrtraceborder}{RGB}{218,221,223}

\newcolumntype{P}[1]{>{\raggedright\arraybackslash}p{#1}}
\newcolumntype{Y}{>{\raggedright\arraybackslash}X}

\newtcolorbox{eagrtracecard}[3][]{%
  enhanced,
  colback=#3,
  colframe=eagrtraceborder,
  colbacktitle=#3,
  coltitle=black!88,
  boxrule=0.35pt,
  arc=0.7mm,
  outer arc=0.7mm,
  boxsep=0pt,
  left=1.5mm,
  right=1.5mm,
  top=0.9mm,
  bottom=0.9mm,
  toptitle=0.9mm,
  bottomtitle=0.6mm,
  before skip=0.8mm,
  after skip=0.8mm,
  title={#2},
  fonttitle=\normalfont\normalsize\bfseries,
  fontupper=\normalfont\normalsize,
  before upper=\raggedright,
  #1
}

\newtcolorbox{eagrtraceoutput}{%
  enhanced,
  colback=white,
  colframe=black!18,
  boxrule=0.35pt,
  arc=0.45mm,
  outer arc=0.45mm,
  boxsep=0pt,
  left=1.3mm,
  right=1.3mm,
  top=0.8mm,
  bottom=0.8mm,
  before skip=0.8mm,
  after skip=0.8mm,
  fontupper=\normalfont\normalsize,
  before upper=\raggedright
}

\newtcolorbox{eagrtracefigure}[1]{%
  enhanced,
  width=\dimexpr\linewidth-1pt\relax,
  center,
  colback=white,
  colframe=black!20,
  boxrule=0.4pt,
  arc=0.6mm,
  outer arc=0.6mm,
  boxsep=0pt,
  left=2mm,
  right=2mm,
  top=4.6mm,
  bottom=1.2mm,
  title={#1},
  halign title=center,
  attach boxed title to top center={yshift=-2.2mm},
  boxed title style={
    colback=black!4,
    colframe=black!18,
    boxrule=0.35pt,
    arc=0.5mm,
    outer arc=0.5mm,
    left=12mm,
    right=12mm,
    top=0.8mm,
    bottom=0.8mm
  },
  fonttitle=\normalfont\normalsize\bfseries,
  coltitle=black!88
}

\newtcblisting{eagrpromptlisting}[2][]{%
  enhanced,
  listing only,
  listing engine=listings,
  colback=eagrtracegray,
  colframe=eagrtraceborder,
  colbacktitle=black!4,
  coltitle=black!88,
  boxrule=0.4pt,
  arc=0.6mm,
  outer arc=0.6mm,
  boxsep=0pt,
  left=1.5mm,
  right=1.5mm,
  top=1mm,
  bottom=1mm,
  toptitle=0.9mm,
  bottomtitle=0.7mm,
  before skip=0.8mm,
  after skip=0.8mm,
  title={#2},
  fonttitle=\normalfont\normalsize\bfseries,
  listing options={
    basicstyle=\ttfamily\normalsize,
    breaklines=true,
    breakatwhitespace=true,
    columns=fullflexible,
    keepspaces=true,
    showstringspaces=false,
    upquote=true,
    tabsize=2,
    aboveskip=0pt,
    belowskip=0pt
  },
  #1
}

\newcommand{\eagroutputkey}{%
  \begin{eagrtracecard}{Joint Node Assessment and Edge Selection Output}{eagrtracegreen}
  \centering
  \textbf{The joint operation returns $(r_i,s_i,K_i',E_i)$ in this fixed order:}

  \smallskip
  \texttt{(is\_relevant, is\_sufficient, [Selected Edge Labels], [Evidence])}
  \end{eagrtracecard}
}

\newcommand{\eagrcompactkey}{%
  \begin{eagrtracecard}{Joint Node Assessment and Edge Selection Output}{eagrtracegreen}
  \textbf{Fixed order:}
  \texttt{(is\_relevant, is\_sufficient, [Selected Edge Labels], [Evidence])}.
  The first two values are Boolean decisions; the third and fourth values are
  lists.
  \end{eagrtracecard}
}

\section{Implicit Graph Construction}
\label{app:graph-construction}

This appendix specifies how descriptors are extracted and normalized and how
neighbourhoods are looked up, so that the offline stage can be reproduced
given the same LLM (temperature~0).

\paragraph{Step 1: Descriptor extraction.}
Each passage, prefixed with its title, is sent to
\textsc{Llama-3.3-70B-Instruct} with the prompt in
Figure~\ref{fig:eagr-extraction-prompt}. The prompt asks for every named
entity and at most five key concepts, as a JSON array of typed items
(\emph{person}, \emph{organization}, \emph{location}, \emph{work},
\emph{event}, \emph{product}, \emph{other}, \emph{concept}, \emph{date}).
Names are requested in their most complete form in the text; partial names
and nationality adjectives are excluded, and no minimum count is imposed.

\paragraph{Step 2: Parsing, cleaning and grounding.}
We parse the first well-formed JSON array in the response. If none is found
(e.g., a missing closing bracket, or a response cut off at the token limit),
the complete entity objects are read individually. Each name is lowercased, hyphens are replaced by spaces, accents are removed
and whitespace is collapsed. A name is kept only if it occurs as a whole-word
phrase in the passage (title included), compared after the same
normalization with punctuation removed; this discards names the LLM invented
or paraphrased. Duplicates are collapsed. Dates are kept in the passage
metadata but excluded from the descriptor set $K_i$.

\paragraph{Step 4: Person-name normalization.}
Using the extracted entity types, we collect person names across the corpus and group them by last name and suffix (e.g., \emph{Jr.}, \emph{II}). Within each group, a name is considered a \emph{shorter form} of another if its words occur in the same order in the longer name, with initials matching words that begin with the corresponding letter (e.g., \emph{Clarence Brown} and \emph{Clarence Leon Brown}; \emph{Robert A. Stemmle} and \emph{Robert Adolf Stemmle}). First names must match, suffixes must be identical, and the shorter name's middle part must match the first middle part of the longer name. Thus, \emph{George W. Bush} is not merged with \emph{George H. W. Bush}. Names containing particles or epithets (e.g., \emph{de}, \emph{van}, \emph{the}, \emph{of}) are excluded from merging. If a shorter form matches at most three full names, all matching full names are added as descriptors to every passage containing that shorter form. If it matches more than three, it is considered ambiguous and left unlinked. Finally, single-word fragments of a person's name (e.g., \emph{Clarence}, \emph{Brown}) are removed from passages containing the full name to prevent spurious connections between different people.

\paragraph{Step 4: Neighborhood lookup.}
At query time, the neighbours along a selected descriptor $k$ are the
passages with at least one descriptor that contains $k$ as a contiguous word
sequence, matched case- and punctuation-insensitively by a phrase query on the
descriptor metadata. A descriptor therefore also reaches passages whose
descriptors extend it (e.g., \emph{the goose woman} matches
\emph{the goose woman remake}). Each lookup
returns at most $N_{\max}$ matching passages per descriptor, without further
ranking among the matches.

\begin{table*}[h]
\centering
\small
\setlength{\tabcolsep}{5pt}
\renewcommand{\arraystretch}{1.15}
\begin{tabularx}{\linewidth}{@{}p{0.11\linewidth} X p{0.11\linewidth} p{0.13\linewidth} p{0.13\linewidth}@{}}
\toprule
\textbf{Category} & \textbf{Question (abridged)} & \textbf{Gold} & \textbf{HippoRAG2} & \textbf{EAGR} \\
\midrule
String false negative &
What three other National Forests are administered by the same people who administer the park the Art Loeb trail runs through? &
Croatan, Nantahala, Uwharrie &
Str$=1$, LLMaJ$=1$ (exact) &
Str$=0$, LLMaJ$=1$ (lists all three with extra text) \\
\addlinespace[2pt]
String false negative &
The Mexican Indignados Movement is a response to a war also known as? &
Mexican War on Drugs &
Str$=1$, LLMaJ$=1$ &
Str$=0$, LLMaJ$=1$ (``Mexican Drug War'') \\
\addlinespace[2pt]
Genuine improvement &
Who directed a 2006 film where Ron Perkins's character plays the manager of a hotel? &
Christopher Nolan &
Str$=0$, LLMaJ$=0$ (``Todd Field'') &
Str$=1$, LLMaJ$=1$ (``Christopher Nolan'') \\
\addlinespace[2pt]
Genuine improvement &
What was the U.S.\ Army position of the Lieutenant General that Robert E.\ Lee surrendered to? &
Commanding General &
Str$=0$, LLMaJ$=0$ (``General-in-Chief'') &
Str$=1$, LLMaJ$=1$ (``Commanding General'') \\
\bottomrule
\end{tabularx}
\caption{Representative HotpotQA cases where String Accuracy and the LLM judge
diverge. The first two rows show EAGR answers that are correct but lexically
penalized; the last two show questions EAGR answers correctly and HippoRAG2
does not.}
\label{tab:hotpot-cases}
\end{table*}

\begin{figure*}[t]
\centering
\begin{minipage}[t]{0.495\textwidth}
\begin{eagrpromptlisting}[colback=eagrtracepeach,
  listing options={basicstyle=\ttfamily\scriptsize, breaklines=true,
                   columns=fullflexible, literate={é}{{\'e}}1}]{Descriptor Extraction Prompt (1/2)}
Extract the entities and key terms from the text
below. They will be used to link this passage to
other passages that mention the same things.

Output ONLY a JSON array, starting with [ and
ending with ]. No explanations, no reasoning, no
markdown.
Each item is an object: {"name": "...", "type": "..."}

Types:
- person: people and characters
- organization: companies, bands, teams,
  institutions, government bodies, political
  parties
- location: countries, cities, regions, rivers
  and other natural features, buildings, landmarks
- work: titles of creative or published works
  (films, songs, albums, books, games, shows,
  artworks, publications). Always include titles,
  including those in quotation marks.
- event: named events, wars, treaties,
  competitions, awards
- product: named products, technologies,
  vehicles, models
- other: any other named entity (for example a
  named chart, list, style or programme)
- concept: specific multi-word terms central to
  the text that other passages would use for the
  same thing. At most 5. Never generic words such
  as "film", "song", "company", "director",
  "album".
- date: dates and years

Rules:
- Include every named entity in the text, however
  many there are. Do not pad the list to reach a
  number.
- Copy each name exactly as written in the text,
  in its most complete form there. Do not invent,
  translate, expand or shorten names.
- If the text itself uses several forms for the
  same entity (a full name and an abbreviation, or
  a title and its translation), output each form
  that appears as a separate item.
\end{eagrpromptlisting}
\end{minipage}\hfill
\begin{minipage}[t]{0.495\textwidth}
\begin{eagrpromptlisting}[colback=eagrtracepeach,
  listing options={basicstyle=\ttfamily\scriptsize, breaklines=true,
                   columns=fullflexible, literate={é}{{\'e}}1}]{Descriptor Extraction Prompt (2/2)}
- Do not output parts of a name on their own,
  unless the text refers to an entity only that
  way.
- Nationality and language adjectives ("British",
  "French") are not entities. Include a country
  only if it is named.
- Pronouns and descriptions ("the singer", "his
  wife") are not entities.
- No duplicates.

Example
Text:
"Northern Lights" is a 1998 song by the Scottish
band Glass Harbour, written by lead singer Mara
Ellison Kent. It appeared on the album "Tidewater"
(French: "Marée"), released by Polar Records, and
reached number 4 on the UK Singles Chart.

Output:
[{"name": "Northern Lights", "type": "work"},
 {"name": "Glass Harbour", "type": "organization"},
 {"name": "Mara Ellison Kent", "type": "person"},
 {"name": "Tidewater", "type": "work"},
 {"name": "Marée", "type": "work"},
 {"name": "Polar Records", "type": "organization"},
 {"name": "UK Singles Chart", "type": "other"},
 {"name": "1998", "type": "date"}]

Text:
{content}

Output:
\end{eagrpromptlisting}
\end{minipage}
\caption{Prompt used in the offline stage to extract typed descriptors for
each passage, shown in two parts. Line breaks are added for layout. The
response is parsed, cleaned, grounded against the passage and normalized as
described in Appendix~\ref{app:graph-construction}.}
\label{fig:eagr-extraction-prompt}
\end{figure*}
\clearpage
\section{Illustrative Retrieval Trace}
\label{app:retrieval-trace}

Figures~\ref{fig:eagr-trace-frontier-zero}--\ref{fig:eagr-trace-frontier-one}
present a complete recorded retrieval trajectory, from seed retrieval to
termination. The organization mirrors the online stages of
\S\ref{sec:eagr}: joint Node Assessment and Edge Selection (one LLM call per
passage), the Sufficiency Check (once per frontier), Evidence-Guided Edge
Reranking and Graph Expansion, followed by Retrieval Reordering. Passages are
abridged to their query-relevant sentences, and long edge-label lists are
shortened; everything else is as recorded.

\begin{figure*}[t!]
\centering
\begin{eagrtracefigure}{Initialization and Frontier 0: Seed 1}

\begin{eagrtracecard}{Initialization: Dense Passage Retrieval (DPR) of Seed Nodes}{eagrtraceblue}
\textbf{Query.} What is the place of birth of Atang de la Rama's husband?

\smallskip
\textbf{Seed retrieval.} Dense retrieval with $k=4$ returns the initial
frontier
$V_0=[v_1\!:\!\textit{Atang de la Rama},\;
v_2\!:\!\textit{Christopher O'Neill},\;
v_3\!:\!\textit{Urata Rama},\;
v_4\!:\!\textit{Renaldo Rama}]$.
The accumulated evidence is initialized as $\mathcal{M}=[\,]$.
\end{eagrtracecard}

\begin{eagrtracecard}{Reading the Joint Output}{eagrtracegray}
\renewcommand{\arraystretch}{1.25}
\begin{tabularx}{\linewidth}{@{}P{0.18\linewidth}P{0.30\linewidth}Y@{}}
\textbf{is\_relevant} ($r_i$) &
\textbf{Selected Edge Labels} ($K_i'$) &
\textbf{Facts} ($E_i$) \\
Does the passage help answer any step of the query? &
Which of the passage's unvisited descriptors should be followed? &
Which facts from the passage fill a step of the query? Grounded facts of
relevant passages are added to $\mathcal{M}$. \\
\end{tabularx}
\end{eagrtracecard}

\begin{eagrtracecard}{Seed 1 of 4 ($v_1$): Atang de la Rama}{eagrtraceblue}
\textbf{Evidence before assessment.} $\mathcal{M}=[\,]$.

\smallskip
\textbf{Current passage (abridged).}
Honorata de la Rama-Hernandez (January 11, 1902 -- July 11, 1991), commonly
known as Atang de la Rama, was a singer and bodabil performer who became the
first Filipina film actress. [\dots] She was married to National Artist for
Literature, Amado V.\ Hernandez.

\smallskip
\textbf{Edge labels (37, abridged).} \textit{amado v.\ hernandez},
\textit{amado vera hernandez}, \textit{atang de la rama}, \textit{pandacan},
\textit{manila}, \textit{dalagang bukid}, \textit{kundiman}, \dots

\begin{eagrtraceoutput}
\textbf{Joint Node Assessment and Edge Selection output}

\smallskip
\renewcommand{\arraystretch}{1.3}
\begin{tabularx}{\linewidth}{@{}P{0.18\linewidth}P{0.30\linewidth}Y@{}}
\textbf{is\_relevant} & \textbf{Selected Edge Labels} & \textbf{Facts} \\
\texttt{True} & [\textit{amado v.\ hernandez}] &
[``Atang de la Rama was married to Amado V.\ Hernandez.''] \\
\end{tabularx}
\end{eagrtraceoutput}

\textbf{Retrieval-state update.} The passage is a relevant bridge: it
identifies the husband. Its fact enters $\mathcal{M}$, and
\textit{amado v.\ hernandez} becomes a candidate edge. Because the passage
also contains another spelling of the same name, \textit{amado vera
hernandez}, that label becomes a candidate too
(Appendix~\ref{app:prompts}), so the husband's passage is reached whichever
spelling it uses.
\end{eagrtracecard}

\end{eagrtracefigure}
\caption{DPR initialization and the joint Node Assessment and Edge Selection
output for the first seed, which identifies the husband.}
\label{fig:eagr-trace-frontier-zero}
\end{figure*}

\begin{figure*}[t!]
\centering
\begin{eagrtracefigure}{Frontier 0: Remaining Seeds and Sufficiency Check}

\begin{eagrtracecard}{Seed 2 of 4 ($v_2$): Christopher O'Neill}{eagrtracepeach}
\textbf{Evidence before assessment.} ``Atang de la Rama was married to Amado
V.\ Hernandez.''

\smallskip
\textbf{Current passage.}
Christopher Paul O'Neill, KNO (born 27 June 1974) is a British-American
financier and husband of Princess Madeleine, Duchess of H\"alsingland and
G\"astrikland, a daughter of King Carl XVI Gustaf of Sweden.

\begin{eagrtraceoutput}
\textbf{Joint Node Assessment and Edge Selection output}

\smallskip
\renewcommand{\arraystretch}{1.3}
\begin{tabularx}{\linewidth}{@{}P{0.18\linewidth}P{0.30\linewidth}Y@{}}
\textbf{is\_relevant} & \textbf{Selected Edge Labels} & \textbf{Facts} \\
\texttt{True} &
[\textit{princess madeleine, duchess of halsingland and gastrikland}] &
[``Christopher O'Neill was born on 27 June 1974.''; ``Christopher O'Neill is
the husband of Princess Madeleine, Duchess of H\"alsingland and
G\"astrikland.''] \\
\end{tabularx}
\end{eagrtraceoutput}

\textbf{Recorded false positive.} The passage describes \emph{a} husband,
but not Atang de la Rama's, so it should have been judged irrelevant. Its
facts and edge enter the evidence and the candidate set. As the next steps
show, the error is contained: the Sufficiency Check does not accept it as an
answer, and reranking places its edge below the husband's.
\end{eagrtracecard}

\begin{eagrtracecard}{Seeds 3 and 4 ($v_3$, $v_4$): Urata Rama, Renaldo Rama}{eagrtracegray}
Two passages retrieved for the shared name \emph{Rama}: a Kosovan sports
shooter and an Albanian footballer. Both are judged irrelevant
(\texttt{is\_relevant = False}), adding no facts or edges.
\end{eagrtracecard}

\begin{eagrtracecard}{Sufficiency Check after Frontier 0}{eagrtracegreen}
\textbf{Accumulated evidence $\mathcal{M}$.}
(1)~``Atang de la Rama was married to Amado V.\ Hernandez.''
(2)~``Christopher O'Neill was born on 27 June 1974.''
(3)~``Christopher O'Neill is the husband of Princess Madeleine, Duchess of
H\"alsingland and G\"astrikland.''

\smallskip
\textbf{Reasoning chain returned by the LLM.}

\begin{center}
\renewcommand{\arraystretch}{1.2}
\begin{tabular}{lll}
\textbf{Entity} & \textbf{Attribute} & \textbf{Value} \\
Atang de la Rama & husband & Amado V.\ Hernandez \\
Amado V.\ Hernandez & place of birth & \textit{(empty)} \\
Christopher O'Neill & place of birth & \textit{(empty)} \\
\end{tabular}
\end{center}

\textbf{Decision.} \texttt{is\_sufficient = No}: the husband's place of birth
is missing from the evidence. Traversal continues.
\end{eagrtracecard}

\end{eagrtracefigure}
\caption{Frontier 0, continued. One seed is a recorded false positive (a
different husband) and two are irrelevant name matches. The Sufficiency Check
identifies the missing step, so traversal continues.}
\label{fig:eagr-trace-seeds}
\end{figure*}

\begin{figure*}[t!]
\centering
\begin{eagrtracefigure}{Reranking, Graph Expansion, Frontier 1 and Termination}

\begin{eagrtracecard}{Evidence-Guided Edge Reranking}{eagrtracegreen}
The candidates selected in frontier 0 are
$\mathcal{K}_{\mathrm{cand}}=$ [\textit{amado v.\ hernandez},
\textit{amado vera hernandez}, \textit{princess madeleine, duchess of
halsingland and gastrikland}]. Given $q$ and $\mathcal{M}$, the reranker
returns
\begin{center}
\textit{amado v.\ hernandez} $\succ$ \textit{princess madeleine, duchess of
halsingland and gastrikland} $\succ$ \textit{amado vera hernandez},
\end{center}
placing the husband's name first.
\end{eagrtracecard}

\begin{eagrtracecard}{Graph Expansion}{eagrtracegreen}
Metadata lookup along $K^*$ reaches one unexplored passage, which forms the
next frontier $V_1=[v_5\!:\!\textit{Amado V.\ Hernandez}]$.
\end{eagrtracecard}

\begin{eagrtracecard}{Frontier 1 ($v_5$): Amado V.\ Hernandez}{eagrtraceblue}
\textbf{Current passage (abridged).}
Amado Vera Hernandez, commonly known as Amado V.\ Hernandez
(September 13, 1903 -- March 24, 1970), was a Filipino writer and labor
leader [\dots]. He was born in Tondo, Manila, to parents from Hagonoy,
Bulacan.

\smallskip
\textbf{Edge labels} (unvisited only). \textit{american correspondence
school}, \textit{bulacan}, \textit{gagalangin}, \textit{hagonoy},
\textit{manila}, \textit{manila high school}, \textit{philippines},
\textit{tondo}

\begin{eagrtraceoutput}
\textbf{Joint Node Assessment and Edge Selection output}

\smallskip
\renewcommand{\arraystretch}{1.3}
\begin{tabularx}{\linewidth}{@{}P{0.18\linewidth}P{0.30\linewidth}Y@{}}
\textbf{is\_relevant} & \textbf{Selected Edge Labels} & \textbf{Facts} \\
\texttt{True} & [\textit{manila}, \textit{bulacan}] &
[``Amado V.\ Hernandez was born in Tondo, Manila.''; ``Amado V.\ Hernandez
was born to parents from Hagonoy, Bulacan.''] \\
\end{tabularx}
\end{eagrtraceoutput}
\end{eagrtracecard}

\begin{eagrtracecard}{Sufficiency Check after Frontier 1: Termination}{eagrtracegreen}
\begin{center}
\renewcommand{\arraystretch}{1.2}
\begin{tabular}{lll}
\textbf{Entity} & \textbf{Attribute} & \textbf{Value} \\
Atang de la Rama & husband & Amado V.\ Hernandez \\
Amado V.\ Hernandez & place of birth & Tondo, Manila \\
Amado V.\ Hernandez & parents' birthplace & Hagonoy, Bulacan \\
\end{tabular}
\end{center}

\textbf{Decision.} \texttt{is\_sufficient = Yes}, and every value appears in
$\mathcal{M}$, so the check succeeds and traversal stops, although the judge
had still selected edges for $v_5$. The chain no longer includes
Christopher O'Neill: the false positive from frontier 0 played no part in
the answer.
\end{eagrtracecard}

\begin{eagrtracecard}{Retrieval Reordering and Cost}{eagrtracegray}
\renewcommand{\arraystretch}{1.25}
\begin{tabularx}{\linewidth}{@{}P{0.06\linewidth}P{0.26\linewidth}P{0.18\linewidth}Y@{}}
\textbf{Rank} & \textbf{Passage} & \textbf{Source} & \textbf{Group} \\
1 & \textit{Atang de la Rama} & Dense seed & (i) supports the chain \\
2 & \textit{Amado V.\ Hernandez} & Graph expansion & (i) supports the chain \\
3 & \textit{Christopher O'Neill} & Dense seed & (ii) relevant, contributed facts \\
4 & \textit{Urata Rama} & Dense seed & (iii) judged irrelevant \\
5 & \textit{Renaldo Rama} & Dense seed & (iii) judged irrelevant \\
\end{tabularx}

\smallskip
Retrieval used 5 of the budget of $B=30$ passages and 8 LLM calls (5 joint
assessments, 2 sufficiency checks and 1 reranking).
\end{eagrtracecard}

\end{eagrtracefigure}
\caption{Reranking, Graph Expansion and Frontier 1. The husband's passage,
reached by one hop, completes the reasoning chain; the Sufficiency Check
succeeds and traversal stops after 5 of 30 passages. Reordering places both
passages of the evidence chain first.}
\label{fig:eagr-trace-frontier-one}
\end{figure*}

\clearpage
\section{Retrieval Prompts}
\label{app:prompts}
\label{app:retrieval-prompts}

EAGR uses three prompts, all sent to the same LLM with greedy decoding and
thinking disabled. The prompts below are reproduced verbatim; only line
breaks are added for layout.

\paragraph{Joint Node Assessment and Edge Selection.}
One call per explored passage (Figures~\ref{fig:eagr-joint-prompt}
and~\ref{fig:eagr-joint-examples}). In the prompt, \emph{Previously Retrieved
Facts} is the accumulated evidence $\mathcal{M}_t$, \emph{Current Passage} is
$p_i$ (prefixed with its title), and \emph{Edge Labels} are the passage's
unvisited descriptors. The output fields \texttt{is\_relevant},
\texttt{facts} and \texttt{edge\_labels} give $r_i$, $E_i$ and $K_i'$; the
\texttt{reasons} field is generated to encourage deliberate selection but is
not used.

\paragraph{Sufficiency Check.}
At most one call per frontier (Figure~\ref{fig:eagr-sufficiency-prompt}). The
prompt receives the query and the full accumulated evidence $\mathcal{M}_t$
and returns the reasoning chain (\texttt{required}) and a judgement
(\texttt{is\_sufficient}), which is verified as described below.

\paragraph{Evidence-Guided Edge Reranking.}
At most one call per frontier, only when the evidence is not yet sufficient
(Figure~\ref{fig:eagr-reranking-prompt}). The prompt receives $q$,
$\mathcal{M}_t$ and $\mathcal{K}_{\mathrm{cand}}$ and returns the ordered
list $K^*$.

\paragraph{Implementation details.}
\begin{itemize}
\item \emph{Candidate edges.} The Edge Labels shown for a passage are its
descriptors that have not yet been expanded. When a person's name is
selected, other spellings of the same name in that passage (e.g.,
\emph{robert a.\ stemmle} and \emph{robert stemmle}) are expanded as well,
since a neighbouring passage may use only one of them.
\item \emph{Parsing.} Selected labels are matched to the candidates ignoring
case and whitespace, and labels that are not among the candidates are
discarded. If the joint prompt's output cannot be parsed, the passage
contributes no evidence or edges, and its relevance is treated as unknown
during reordering.
\item \emph{Fact grounding.} Since the LLM may introduce information that is
not in the passage, a fact is kept only if every number in it appears in the
passage and at least 60\% of its content words do (words of five or more
letters may match on their first five letters, so that, e.g.,
\emph{married} matches \emph{marriage}). A fact is skipped if an existing
fact in $\mathcal{M}_t$ already states it, i.e., contains its numbers and at
least 85\% of its content words.
\item \emph{Sufficiency verification.} $s_t=1$ only if the LLM answers
\texttt{Yes}, the chain has at least one step, and every value (every item
of a list value) appears in $\mathcal{M}_t$: its numbers must match exactly,
and either the whole value or all of its content words must occur.
\item \emph{Reranking fallback.} Labels missing from the reranker's output are
appended in their selection order; if the output cannot be parsed, the
selection order is kept.
\item \emph{Evidence chain for reordering.} A passage supports the reasoning
chain if one of its facts contains a chain entity or value as a phrase, or
contains all its numbers and at least 75\% of its content words.
\end{itemize}

\begin{figure*}[p]
\centering
\begin{minipage}[t]{0.495\textwidth}
\begin{eagrpromptlisting}[colback=eagrtraceblue, eagrsmall]{Joint Prompt: Instructions (1/2)}
You are helping a multi-hop retrieval system answer a query by moving from passage to passage.

You are given a query, the facts retrieved so far, the CURRENT passage, and the edge labels of the current passage. Following an edge label retrieves other passages that mention that label. The list of edge labels may be empty.

Fill in these fields:

1. is_relevant: "Yes" if the passage helps answer any step of the query: it gives information the query asks for, or it identifies an entity on the path to the answer (for example the person, place, organization or work that a query entity points to). The passage does not need to contain the final answer. "No" if the passage has no connection to the query, or if it is about a different entity that merely has the same role or the same kind of information as an entity the query needs (for example another person with the same occupation, or another work with a similar title).
2. facts: if is_relevant is "Yes", list ONLY the facts from the current passage that fill a step of the query: facts that state what the query asks about, or facts that identify the next entity in the chain. Leave out facts that are true but would not change the answer, such as other roles, other works, or other events involving the same entities. For queries that compare entities or ask what they have in common, include the values or candidates needed for each compared side. Write each fact as a short, self-contained sentence (use names, not pronouns). Use only the current passage: do not copy facts from the previously retrieved facts or the query, and do not add anything from your own knowledge. If is_relevant is "No", return [].
3. edge_labels: if is_relevant is "Yes", select the edge labels that are likely to lead to NEW passages containing information still needed to answer the query. If is_relevant is "No", return [].
4. reasons: one brief reason for each selected edge label, in the same order.
\end{eagrpromptlisting}
\end{minipage}\hfill
\begin{minipage}[t]{0.495\textwidth}
\begin{eagrpromptlisting}[colback=eagrtraceblue, eagrsmall]{Joint Prompt: Instructions (2/2)}
Rules for selecting edge labels:
- Select a label only if following it from the CURRENT passage is likely to uncover NEW information needed for the query. Ask: "If I follow this label, will I find information I still need?" If not, do not select it.
- Do not select a label just because it appears in the passage or in the query.
- Do not select broad categories or generic terms unless they are the only path to information still needed.
- Do not select labels that only match the answer type without leading to the answer.
- Do not select labels that would only retrieve information already present in the previously retrieved facts.
- Copy labels exactly as they appear in the Edge Labels list. Never invent, modify or normalize a label.
- Any number of labels from zero to all of them may be selected. Do not force a selection.

The previously retrieved facts are context only. They are not a retrieval source: never select labels or extract facts from them.

Return ONLY one JSON object with exactly these keys and no other text:
{"is_relevant": "Yes" or "No", "facts": [...], "edge_labels": [...], "reasons": [...]}

[Four examples: Figure B.2]

Now the actual input.
Query:
{query}

Previously Retrieved Facts (read-only, never expand):
{accumulated evidence, or None}

Current Passage:
{passage title and text}

Edge Labels:
{JSON list of unvisited descriptors}

Output:
\end{eagrpromptlisting}
\end{minipage}
\caption{Joint Node Assessment and Edge Selection prompt, with its runtime
input template. The few-shot examples, which appear between the instructions
and the input, are shown in Figure~\ref{fig:eagr-joint-examples}.}
\label{fig:eagr-joint-prompt}
\end{figure*}

\begin{figure*}[p]
\centering
\begin{minipage}[t]{0.495\textwidth}
\begin{eagrpromptlisting}[colback=eagrtracepeach, eagrsmall]{Joint Prompt: Examples 1 and 2}
Example 1
Query:
What city was the inventor of the device created by Sarah Chen born in?

Previously Retrieved Facts (read-only, never expand):
None

Current Passage:
Sarah Chen founded Lumina Robotics in 2012 and led the development of the SmartGuide navigation device. The device was invented by engineer Michael Torres and later adopted by several hospitals.

Edge Labels:
["sarah chen", "lumina robotics", "smartguide navigation device", "michael torres", "engineer", "hospitals", "inventor", "founder"]

Output:
{"is_relevant": "Yes", "facts": ["Sarah Chen led the development of the SmartGuide navigation device.", "The SmartGuide navigation device was invented by engineer Michael Torres."], "edge_labels": ["smartguide navigation device", "michael torres"], "reasons": ["Passages about the SmartGuide navigation device may describe its inventor.", "Passages about Michael Torres, the inventor, may give his birthplace."]}


Example 2
Query:
What city was the inventor of the device created by Sarah Chen born in?

Previously Retrieved Facts (read-only, never expand):
Sarah Chen led the development of the SmartGuide navigation device. The SmartGuide navigation device was invented by engineer Michael Torres.

Current Passage:
Michael Torres (born April 18, 1982, in Portland, Oregon, USA) is a civil engineer and environmental consultant specializing in sustainable urban infrastructure.

Edge Labels:
["michael torres", "portland", "oregon", "civil engineer", "environmental consultant"]

Output:
{"is_relevant": "Yes", "facts": ["Michael Torres was born in Portland, Oregon, USA."], "edge_labels": [], "reasons": []}
\end{eagrpromptlisting}
\end{minipage}\hfill
\begin{minipage}[t]{0.495\textwidth}
\begin{eagrpromptlisting}[colback=eagrtracepeach, eagrsmall]{Joint Prompt: Examples 3 and 4}
Example 3
Query:
When did the director of the film Blue Lantern die?

Previously Retrieved Facts (read-only, never expand):
Blue Lantern was directed by Henry Walsh.

Current Passage:
Arthur Bennett (2 March 1910 – 14 June 1985) was an English film director and screenwriter.

Edge Labels:
["arthur bennett", "english", "film director", "screenwriter"]

Output:
{"is_relevant": "No", "facts": [], "edge_labels": [], "reasons": []}


Example 4
Query:
Which scientist was born later, the inventor of the Aurora Engine or the inventor of the Horizon Drive?

Previously Retrieved Facts (read-only, never expand):
The Aurora Engine was invented by Emma Carter in 1998. The Horizon Drive was invented by Daniel Brooks.

Current Passage:
Daniel Brooks (born March 3, 1975) is an American aerospace engineer.

Edge Labels:
[]

Output:
{"is_relevant": "Yes", "facts": ["Daniel Brooks was born on March 3, 1975."], "edge_labels": [], "reasons": []}
\end{eagrpromptlisting}
\end{minipage}
\caption{Few-shot examples of the joint prompt: (1)~a relevant passage that
identifies the next entity, with two selected edges; (2)~a relevant passage
that completes the evidence; (3)~a passage about a different entity in the
same role, judged irrelevant; (4)~a relevant passage with no candidate edges,
whose fact is still added to the evidence.}
\label{fig:eagr-joint-examples}
\end{figure*}

\begin{figure*}[p]
\centering
\begin{minipage}[t]{0.495\textwidth}
\begin{eagrpromptlisting}[colback=eagrtracegray, eagrsmall]{Sufficiency Check Prompt}
You decide whether a set of retrieved facts is enough to answer a query.

1. required: write out the steps needed to answer the query, in order. Each step has:
   - "entity": what the step is about. Start from the entities in the query; once a step finds an entity (for example the director), name it in later steps ("Claire Dubois", not "the director").
   - "attribute": what is needed about that entity.
   - "value": the value as written in the facts, or "" if the facts do not state it. If a step has several values, give them as a list.
   For comparisons and yes/no questions, include each compared value as its own step (for example the birth date of each person); the answer follows from those values.
   Use ONLY the facts below, never your own knowledge, and never guess. The value must be stated in the facts about that entity, wherever it appears in a sentence: "Paul Ward, the brother of Rivertown Hawks radio broadcaster Sam Ward" states that Sam Ward broadcast for the Rivertown Hawks, but "a French film directed by Claire Dubois" does not state Claire Dubois's nationality. Answer the attribute the query asks for, not a different one the facts happen to contain.
2. is_sufficient: "Yes" only if every value is filled in. Otherwise "No".

Return ONLY one JSON object, with no other text:
{"required": [{"entity": "...", "attribute": "...", "value": "..."}, ...], "is_sufficient": "Yes" or "No"}

[Two examples: right]

Now the actual input.
Query:
{query}

Facts:
{accumulated evidence}

Output:
\end{eagrpromptlisting}
\end{minipage}\hfill
\begin{minipage}[t]{0.495\textwidth}
\begin{eagrpromptlisting}[colback=eagrtracepeach, eagrsmall]{Sufficiency Check: Examples}
Example 1
Query:
Are the directors of Silent Harbor and The Glass Orchard from the same country?

Facts:
Silent Harbor is a 1998 French drama film directed by Claire Dubois. The Glass Orchard is a 2004 film directed by Marco Rossi. Marco Rossi is an Italian film director.

Output:
{"required": [{"entity": "Silent Harbor", "attribute": "director", "value": "Claire Dubois"}, {"entity": "The Glass Orchard", "attribute": "director", "value": "Marco Rossi"}, {"entity": "Claire Dubois", "attribute": "nationality", "value": ""}, {"entity": "Marco Rossi", "attribute": "nationality", "value": "Italian"}], "is_sufficient": "No"}


Example 2
Query:
What city was the inventor of the device created by Sarah Chen born in?

Facts:
Sarah Chen led the development of the SmartGuide navigation device. The SmartGuide navigation device was invented by engineer Michael Torres. Michael Torres was born in Portland, Oregon, USA.

Output:
{"required": [{"entity": "Sarah Chen", "attribute": "device created", "value": "SmartGuide navigation device"}, {"entity": "SmartGuide navigation device", "attribute": "inventor", "value": "Michael Torres"}, {"entity": "Michael Torres", "attribute": "birthplace", "value": "Portland, Oregon, USA"}], "is_sufficient": "Yes"}
\end{eagrpromptlisting}
\end{minipage}
\caption{Sufficiency Check prompt, with its runtime input template and
few-shot examples: (1)~insufficient, since one director's nationality is not
stated (the film's country does not count), and (2)~sufficient, since every
step of the chain is filled from the facts.}
\label{fig:eagr-sufficiency-prompt}
\end{figure*}

\begin{figure*}[t]
\centering
\begin{eagrpromptlisting}[colback=eagrtracegreen, eagrsmall]{Evidence-Guided Edge Reranking Prompt}
You are a retrieval assistant. Given a query, previously retrieved facts, and a list of candidate edge labels, rank the edge labels from most to least useful for retrieving passages that help answer the query.

Instructions:
- Return ONLY a JSON array of the edge labels in ranked order, most useful first.
- Include every label from the list exactly as written. Do not add, remove or modify any label.
- Do not include explanations, markdown or any text outside the JSON array.

Query:
{query}

Previously Retrieved Facts:
{accumulated evidence, or None}

Candidate Edge Labels:
{JSON list of the descriptors selected in the current frontier}

Output:
\end{eagrpromptlisting}
\caption{Evidence-Guided Edge Reranking prompt, which orders the candidate
set $\mathcal{K}_{\mathrm{cand}}$ into $K^*$.}
\label{fig:eagr-reranking-prompt}
\end{figure*}

\end{document}